\documentclass[prd,twocolumn,floats,floatfix, nofootinbib,center]{revtex4}
\usepackage[utf8]{inputenc}
\usepackage{color}

\usepackage{graphicx}
\usepackage{mathtools, makecell}

\usepackage{longtable}

\usepackage{hyperref}
\usepackage{cleveref} 

\newcommand{\scriptR}{\mathcal{R}}

\begin{document}

\title{Bouncing cosmology from nonlinear dark energy with two cosmological constants}

\author{Molly Burkmar${{^a}}$\footnote{molly.burkmar@port.ac.uk} and Marco Bruni${{^{a,b}}}$\footnote{marco.bruni@port.ac.uk} 
\vspace{0.5cm}}

\affiliation{${{^a}}$Institute of Cosmology {\rm \&} Gravitation, University of Portsmouth, Dennis Sciama Building, Burnaby Road, Portsmouth, PO1 3FX, United Kingdom\\
${{^b}}$INFN Sezione di Trieste, Via Valerio 2, 34127 Trieste, Italy
}
\date{\today}

\begin{abstract}
    We explore the dynamics of FLRW cosmologies which consist of dark matter, radiation and dark energy with a quadratic equation of state. Standard cosmological singularities arise due to energy conditions which are violated by dark energy, therefore we focus our analysis on non-singular bouncing and cyclic cosmologies, in particular focusing on the possibility of closed FLRW models always having a bounce for any initial conditions. We analyse the range of dynamical behaviour admitted by the system, and find a class of closed models that admit a non-singular bounce, with early- and late-time accelerated expansion connected by a decelerating phase. In all cases, we find the bouncing models are only relevant when dark matter and radiation appear at a certain energy scale, and so require a period such as reheating. We then investigate imposing an upper bound on the dark matter and radiation, such that their energy densities cannot become infinite. In this case, we find that all closed models bounce, and a class of models exist with early- and late-time acceleration, connected by a decelerating phase. We also consider parameter values for the dark energy component, such that the discrepancy between the observed value of $\Lambda$ and the theoretical estimates of the contributions to the effective cosmological constant expected from quantum field theory would be explained. However, we find that the class of models left does not allow for an early- and late-time accelerated expansion, connected by a decelerating period where large-scale structure could form. Nonetheless, our qualitative analysis serves as a basis for the construction of more realistic models with realistic quantitative behaviour.
\end{abstract}

\maketitle

\twocolumngrid

\section{Introduction}

The $\Lambda$-cold-dark-matter ($\Lambda$CDM) model has provided a successful framework to describe the history of the Universe, and is consistent with a wealth of observational data \cite{Planck2018I, Planck2018VI}. Despite its robustness, there are problems which require addressing. The issue of the inevitability of a singularity as the origin of our Universe has been debated for many years. Assuming the strong energy condition holds, singularities appear to be unavoidable \cite{Penrose1965,Hawking1965,Hawking1966I, Hawking1966II, Hawking1967III,Hawking1970, Hawking1973, Wald1984}, however their current interpretation is that they represent points where General Relativity breaks down \cite{Joshi1993,Hawking1979,Gibbons2003,Ashtekar2015}. Broadly speaking, there are two ways to tackle this problem: solve the singularity problem by developing a modified theory of gravity to unify General Relativity with this high energy regime, or avoid the singularity by breaking the standard energy conditions in an alternative origin story of the Universe in General Relativity. For the latter, nonsingular bouncing models have been proposed as a way to evade a singularity \cite{Gasperini1993,Khoury2001,Kallosh2001,Steinhardt2002,Novello2008,Wands2008,Maier2012,Maier2013,Battefeld2014, Cai2017i,Brandenberger2017,Cai2017ii,Ijjas2018,Ijjas2019,Ganguly2019,Ilyas2020,Battista2021,Zhu2021}.

Another problem facing $\Lambda$CDM is that of $\Lambda$ itself. Observations have provided strong evidence for the accelerated expansion of our Universe \cite{Riess1998,Perlmutter1998}, which requires a component that violates the strong energy condition \cite{Hawking1973}, known as dark energy. 
The cosmological constant $\Lambda$ is the simplest explanation we have for dark energy, however the observed value of $\Lambda$ is at odds by up to 120 orders of magnitude with theoretical estimates of the contributions to the effective cosmological constant expected from quantum field theory (QFT) \cite{Weinberg1989, Straumann1999, Ellis2012}.

Taking into account a dynamic dark energy component beyond $\Lambda$, it is worth considering whether a bounce can be produced. A non-interacting vacuum is equivalent to $\Lambda$ in General Relativity \cite{Lemaitre1934}, so a simple extension is to consider a vacuum that interacts with other components, thereby becoming dynamical \cite{Wands2012}. It has been shown that non-singular bounces can arise for both a linear and non-linear interaction of a vacuum with a perfect fluid in Friedmann-Lema{\^i}tre-Robertson-Walker (FLRW) cosmologies \cite{Bruni2022}. 

For non-interacting models, dark energy represented by a single fluid with a quadratic equation of state (EoS) in FLRW can admit nonsingular bounces with the right combination of parameters \cite{Ananda2006}. In addition, a quadratic EoS finds motivations from e.g.\ brane-world models \cite{Shiromizu2000,Bridgman2002,Gergely2002,Langlois2003,Maartens2010}, loop quantum cosmology \cite{Vandersloot2005, Parisi2007} and k-essence \cite{Scherrer2004,Giannakis2005} (see also \cite{Ananda2006,Ananda20062} and references therein). 
 In general, a nonlinear EoS can allow for the existence of effective cosmological constants, appearing as asymptotic states of the dynamical dark energy. Although simple, analysing a quadratic EoS serves well to illustrate the general qualitative behaviour of a system that can admit two effective cosmological constants, a high energy one acting as a repellor and a low energy one acting as a future attractor for the expanding models. Therefore, it is useful to understand the global dynamics as it is only dependent on the existence of the effective cosmological constants and not on the specific model used to produce them. 

In this paper, we analyse a subset of the quadratic EoS from \cite{Ananda2006,Ananda20062} which admit two effective cosmological constants and produce a bounce for models with positively curved spatial sections, i.e.\ closed FLRW models. This dark energy only violates the strong energy condition, and not the null energy condition, hence a bounce can only occur when the curvature is positive. Here, we extend the scenario in \cite{Ananda2006,Ananda20062} to include non-interacting dark matter and radiation in order to understand the effect these components have on the bounce. In particular, we investigate whether all closed models will bounce when matter and radiation are included, thereby providing a scenario where the bounce is generic, i.e. does not depend on initial conditions. Standard radiation and dark matter diverge more quickly at high-enough energies than the curvature and dark energy considered here. We find that we obtain bouncing models for certain values of parameters, but at high energies, if dark matter and radiation become dominant, some models become singular. We therefore consider models where dark matter and radiation are produced at a certain energy scale after the bounce, by imposing an upper bound on their energy densities. In this case, we find that all closed models bounce.

We would also like to understand whether such a model could alleviate the cosmological constant problem, and unite theoretical estimates from QFT with observational constraints on $\Lambda$. We restrict our analysis such that the dark energy has an effective EoS parameter satisfying $w > -1$, so that it decays during the expansion between the high energy and low energy cosmological constants, even allowing a decelerated phase in general, and do not consider phantom dark energy ($w < -1$). The analysis for this paper is available through GitHub, provided the reader has a Mathematica license \cite{BurkmarNotebook}.

The paper is organized as follows. In Section \ref{sec:Dynamical_Eqns} we present the system of equations we are analysing in terms of dimensionless variables as well as in compactified form. In Section \ref{sec:Submanifolds} we show the sub-manifolds of the system, where the dimensions of the dynamics is reduced and we explore the stability of the fixed points. The sub-manifolds of the system include cases analogous to $\Lambda$CDM, and ones in which only dark energy is present which highlight the dynamics before including dark matter and radiation. In Section \ref{sec:Full_System} we study the dynamics of the full system, where both dark matter and radiation are present in standard form, i.e. with EoS parameters $w_m = 0$ and $w_r = 1/3$. We show that a class of closed models exist where bounces are obtained with early- and late-time accelerated expansion phases, connected by a decelerating phase. The caveat with this full system is that not all closed models bounce, and above a certain energy scale matter and radiation become dominant, and, depending on the initial conditions, a subset of closed models evolve between two singularities. Therefore, these models require a post-bounce process such as reheating, when matter and radiation are created. In this spirit, in Section \ref{sec:FullSystemUpperBound} we re-consider the equations for dark matter and radiation, introducing scale-dependent EoS's that limit their energy densities to an upper bound, i.e. an energy scale at which they appear, such that at lower energies we still have $w_m \rightarrow 0$ and $w_r \rightarrow 1/3$. We present the phase spaces for the new system of equations, showing that in general all closed models bounce. We find that it is possible for these models to evolve with early- and late-time accelerated expansion separated by a deceleration period, however the cosmological constant problem cannot be solved simultaneously. We conclude in Section \ref{sec:Conclusions}. In this paper, we employ natural units such that $8 \pi G = c = 1$.

\section{Cosmological Dynamical System} \label{sec:Dynamical_Eqns}

\subsection{Cosmology with nonlinear EoS dark energy}

FLRW models with no cosmological constant term in Einstein's field equations evolve according to a dynamical system, consisting of the energy conservation equations for each component and the Raychaudhuri equation. We consider the dynamics of a universe filled with dark energy, pressureless dark matter and radiation, which evolve according to

\begin{equation}
    \dot{\rho}_x=-3H(\rho_x+P_x)
    \label{rhoxdot}
\end{equation}

\begin{equation}
    \dot{\rho}_m=-3H\rho_m
    \label{rhomdot}
\end{equation}

\begin{equation}
    \dot{\rho}_r=-4H\rho_r\,,
    \label{rhordot}
\end{equation}

\noindent where $\rho_x$ and $P_x$ are the energy density and isotropic pressure of dark energy, $\rho_m$ is the energy density of dark matter and $\rho_r$ is the energy density of radiation; overdots are derivatives with respect to time $t$. $H$ is the Hubble expansion function, and the Raychaudhuri equation describing it's evolution is

\begin{equation}
    \dot{H}=-H^2-\frac{1}{6}(\rho_x + 3P_x + \rho_m + 2\rho_r)\,.
    \label{Hubble_rate}
\end{equation}

We assume that the dark energy EoS is barotropic, $P_x = P_x(\rho_x)$, with $P_x(\rho_x) \geq -\rho_x$, such that Eq. \eqref{rhoxdot} has two fixed points ($\dot{\rho}_x=0$), corresponding to two effective cosmological constants: 
\begin{equation}\label{ecc}
    P_x(\rho_\Lambda) = -\rho_\Lambda, ~~~ P_x(\rho_*) = -\rho_*.
\end{equation}
\noindent 

It follows from the assumption $P_x(\rho_x) \geq -\rho_x$ that 
the dark energy density decreases/increases during a period of expansion/contraction. 
Then, during expansion ($H>0$),  we take $\rho_\Lambda$ to represent the  attractor at low energy close to the dark energy density we observe today, and  $\rho_*$ the repellor at high energy; their roles are inverted during contraction. We assume $\rho_*$ is positive and is an energy scale between the Planck scale and that typical of inflation, to ensure the evolution is always classical, but does not interfere with particle production at lower energies. 

In this paper we assume the same quadratic EoS used in \cite{Ananda2006,Ananda20062}, which is the simplest non-linear case to study and, as we said in the introduction (see also \cite{Ananda2006,Ananda20062}),  has other motivations from various scenarios \cite{Shiromizu2000,Bridgman2002,Gergely2002,Langlois2003,Maartens2010,Vandersloot2005, Parisi2007,Scherrer2004,Giannakis2005}. With respect to \cite{Ananda2006,Ananda20062}, we restrict parameters such that the linear term is always positive, $\alpha > 0$, the quadratic term is always negative, $\beta = -1/\rho_*$, and the constant pressure term is negative, $P_0 = -\rho_\Lambda$. With this choice of parameters our EoS is

\begin{equation}
    P_x=-\rho_\Lambda+\frac{\rho_\Lambda}{\rho_*}\rho_x-\frac{\rho_x^2}{\rho_*}\,.
\label{eqn:DE_EoS}
\end{equation}

This secures the two effective cosmological constants in Eq. \eqref{ecc},
as now the energy conservation equation of the dark energy \eqref{rhoxdot} can be written as

\begin{equation}
    \dot{\rho_x}=-3H(\rho_x-\rho_\Lambda)\left(1-\frac{\rho_x}{\rho_*}\right).
    \label{rhoxdotEoS}
\end{equation}

We note however that the qualitative dynamics that follows from the the specific EoS \eqref{eqn:DE_EoS} is going to be representative of the dynamics for any EoS satisfying the condition $P_x(\rho_x) + \rho_x > 0$ for $\rho_\Lambda < \rho_x < \rho_*$, i.e. between the two fixed points \eqref{ecc} satisfying $P_x(\rho_x) = -\rho_x$, as these conditions are enough to describe any monotonically decreasing $\rho_x$ between $\rho_*$ and $\rho_\Lambda$ during expansion, see e.g.\ the EoS in \cite{Hogg2021}.

The effective EoS parameter $w_x = P_x/\rho_x$ for the dark energy is as

\begin{equation}
    w_x = -\frac{\rho_\Lambda}{\rho_x} + \frac{\rho_\Lambda}{\rho_*} - \frac{\rho_x}{\rho_*}\,,
    \label{eqn:EoS_DE}
\end{equation}

\noindent and the EoS parameters for dark matter and radiation are $w_m = 0$ and $w_r = 1/3$, respectively. This system admits a first integral, the Friedmann equation, which is written in terms of each energy density as

\begin{equation}
    H^2=\frac{\rho_m}{3}+\frac{\rho_r}{3}+\frac{\rho_x}{3}-\frac{k}{a^2}\,,
    \label{eqn:Friedmann_Dimensions}
\end{equation}

\noindent where $k$ is the curvature and $a$ is the cosmic scale factor, connected to the Hubble expansion function through the expression $H = \dot{a}/a$. We set $a_0 = 1$ today, therefore $k$ is an arbitrary constant which is positive, negative or zero for closed, open and flat models, respectively.

\subsection{Dimensionless Variables} \label{sec:Dimensionless_Variables}

Examining the equations above, a dimensional analysis suggests that the dynamics really only depends on a single parameter, the dimensionless ratio of the two effective cosmological constants, if we use  dimensionless variables.  Following \cite{Ananda2006,Ananda20062}, we define these variables as:

\begin{equation}
    x=\frac{\rho_x}{\rho_*} ~~ y=\frac{H}{\sqrt{\rho_*}} ~~ z=\frac{\rho_m}{\rho_*} ~~
    r=\frac{\rho_r}{\rho_*} ~~ \scriptR=\frac{\rho_\Lambda}{\rho_*} ~~ \eta=\sqrt{\rho_*}t.
    \label{eqn:dimensionless_variables}
\end{equation}

\noindent The variable $\scriptR$ is the ratio of the low energy effective cosmological constant $\rho_\Lambda$ to the high energy effective cosmological constant $\rho_*$ and takes values in the range $(0,1)$,  $x$ is the normalised dark energy density, varying between the two dimensionless effective cosmological constants in the range $[\scriptR,1]$.  The variable $y$ is the normalised Hubble parameter, $z$ the normalised dark matter energy density, $r$ the normalised radiation energy density and $\eta$ the normalised time variable. 
We consider the region of phase space where the energy densities for matter and radiation are always positive, i.e. $z, r > 0$.  
Equations \eqref{rhomdot}, \eqref{rhordot} and \eqref{rhoxdotEoS} then become

\begin{equation}
    z'=-3yz
    \label{darkmatterDE}
\end{equation}

\begin{equation}
    r'=-4yr
    \label{radiationDE}
\end{equation}

\begin{equation}
    x'=-3y(x-\scriptR)(1-x)\,,
    \label{darkenergyDE}
\end{equation}

\noindent where the primes indicate differentiation with respect to $\eta$. The Raychaudhuri equation \eqref{Hubble_rate} can now be expressed as

\begin{equation}
    y'=-y^2-\frac{1}{6}[z+2r-3\scriptR+(1+3\scriptR)x-3x^2]\,,
    \label{HubbleDE}
\end{equation}

\noindent and the Friedmann equation \eqref{eqn:Friedmann_Dimensions} becomes

\begin{equation}
    y^2=\frac{x}{3}+\frac{z}{3}+\frac{r}{3}-\frac{k}{\rho_* a^2}.
    \label{eqn:Friedmann_dimensionless}
\end{equation}

\noindent The effective EoS parameter for the dark energy \eqref{eqn:EoS_DE} in dimensionless variables becomes

\begin{equation}
    w_x = -\frac{\scriptR}{x} + \scriptR - x,
    \label{eqn:EoS_x}
\end{equation}

\noindent therefore, the necessary condition for acceleration is

\begin{equation}
    -\frac{\scriptR}{x} + \scriptR - x < -1/3.
    \label{eqn:accn_condition}
\end{equation}

\noindent We can also express the evolution of the cosmic scale factor $a$ using our dimensionless variables

\begin{equation}
    a'=ay\,.
    \label{eqn:a'}
\end{equation}

\noindent The Friedmann equation \eqref{eqn:Friedmann_dimensionless} can then be rearranged for a kinetic term $a'^2/2$, a potential $U$ and total energy $E$

\begin{equation}
    \frac{a'^2}{2} + U = E\,,
\end{equation}

\noindent where $U$ is given by

\begin{equation}
    U = -\frac{a^2}{6}(x + z + r)
    \label{eqn:U}
\end{equation}

\noindent and $E$ by

\begin{equation}
    E = -\frac{k}{2\rho_*}\,.
    \label{eqn:E}
\end{equation}

\noindent $E$ is zero, positive and negative for flat, open and closed models, respectively.

From Eq. \eqref{eqn:a'} and Eq.s \eqref{darkmatterDE} and \eqref{radiationDE}, we get the standard behaviour for matter and radiation, $z(a) \sim a^{-3}$ and $r(a) \sim a^{-4}$, respectively. For $x$ \eqref{darkenergyDE}, we obtain

\begin{equation}
    x(a) = \frac{a^{-3(1-\scriptR)}+c_a\scriptR}{a^{-3(1-\scriptR)}+c_a}\,,
    \label{eqn:x(a)}
\end{equation}

\noindent where the constant of integration is

\begin{equation}
    c_a=\frac{1-x_0}{x_0-\scriptR}\,.
    \label{eqn:ca}
\end{equation}

\noindent In the following, however, we will express $z$, $r$ and $a$ as functions of $x$.

\subsection{Compactified Variables} \label{sec:Compact_Variables}

In our system, $y$ takes values in the range $(-\infty, \infty)$, and $z$ and $r$ in the range $[0, \infty)$. $x$ is already limited to the range $[\scriptR, 1]$. In order to analyse the full dynamics, it is desirable to deal with a compact phase space. Therefore, we compactify the $y$, $z$ and $r$ variables in the following way:

\begin{equation}
    Y = \frac{y}{\sqrt{1+y^2}}
\end{equation}

\begin{equation}
    Z = \frac{z}{1+z}
\end{equation}

\begin{equation}
    R = \frac{r}{1+r}\,,
\end{equation}

\noindent 
so that $Y$ takes values in the range $[-1,1]$ and  $R$ and $Z$ in the range $[0,1]$. Using these variables, Eq. \eqref{darkmatterDE} - \eqref{HubbleDE} become

\begin{equation}
    Z' = \frac{-3YZ(1 - Z)}{\sqrt{1-Y^2}}
    \label{CompactDarkMatterDE}
\end{equation}

\begin{equation}
    R' = \frac{-4YR(1 - R)}{\sqrt{1-Y^2}}
    \label{CompactRadiationDE}
\end{equation}

\begin{equation}
    x' = \frac{-3Y(x-\scriptR)(1-x)}{\sqrt{1-Y^2}}
    \label{CompactDarkEnergyDE}
\end{equation}

\begin{multline}
    Y' = -Y^2\sqrt{1-Y^2} - \frac{(1-Y^2)^\frac{3}{2}}{6} \\ \times\left[\frac{Z}{1-Z} + \frac{2R}{1-R} + x(1 + 3\scriptR) - 3\scriptR -3x^2 \right]\,,
    \label{CompactHubbleRateDE}
\end{multline}



\noindent and the Friedmann equation \eqref{eqn:Friedmann_dimensionless} can be expressed as

\begin{equation}
    \frac{Y^2}{1-Y^2}=\frac{x}{3}+\frac{Z}{3(1-Z)}+\frac{R}{3(1-R)}-\frac{k}{\rho_* a^2}\,.
    \label{eqn:Friedmann_Dimensionless_Compact}
\end{equation}

\noindent Now, setting $a_0 = 1$, we find

\begin{equation}
    \frac{k}{\rho_*} = \frac{x_0}{3}+\frac{Z_0}{3(1-Z_0)}+\frac{R_0}{3(1-R_0)} - \frac{Y_0^2}{1-Y_0^2}\,.
    \label{eqn:k/rho}
\end{equation}

\noindent Finally, the potential $U$ \eqref{eqn:U} in terms of compact variables is given by

\begin{equation}
    U = -\frac{a^2}{6}\left(x + \frac{Z}{1-Z} + \frac{R}{1-R}\right)\,.
    \label{eqn:compact_U}
\end{equation}

\section{Sub-manifolds of the System} \label{sec:Submanifolds}

To analyse our autonomous dynamical system, we first need to find the sub-manifolds and the fixed points, and determine their linear stability character \cite{Arrowsmith1982, Arrowsmith1992, Strogatz2015}. The fixed points $u_j^*$ satisfy $f_i(u_j^*) = 0$, where $f_i$ are the first-order derivatives of the independent variables $u_j$ with respect to time. We then linearize about each fixed point, first finding the Jacobian matrix, which has the form

\begin{equation}
    \textbf{M}_{ij} = \frac{\partial f_i}{\partial u_j}.
\end{equation}

\newpage
\onecolumngrid

\begingroup
\LTcapwidth=\textwidth
\squeezetable

\begin{longtable}{|| c | c | c | c | c ||}

 \hline
 Sub-manifold & Fixed Points & Name & Eigenvalues & Stability Character \\ [0.5ex] 
 \hline
 \endfirsthead

 \hline
 Sub-manifold & Fixed Points & Name & Eigenvalues & Stability Character \\ [0.5ex] 
 \hline
 \endhead

 \hline
 \endfoot

 \endlastfoot

 \hline\hline
 $x = 1$, $Z = 0$ & $Y = \pm \frac{1}{2}$, $R = 0$ & $dS_{1\pm}$ & $\begin{pmatrix}
\mp \frac{4}{\sqrt{3}}\\
\mp \frac{2}{\sqrt{3}} \\
\end{pmatrix}$ & Attractor/Repellor \\
 & $Y = 0$, $R = \frac{1}{2}$ & $E$ &
 $\begin{pmatrix}
\frac{2}{\sqrt{3}}\\
-\frac{2}{\sqrt{3}} 
\end{pmatrix}$  & Saddle \\
& $Y = \pm 1$, $R = 0$ & $dS_{2\pm}$ & N/A & Saddle\\
& $Y = \pm 1$, $R = 1$ & $S_\pm$ & N/A & Repellor/Attractor \\
\hline
 $x = 1$, $R = 0$ & $Y = \pm \frac{1}{2}$, $Z = 0$ & $dS_{1\pm}$ &$\begin{pmatrix}
\mp \sqrt{3} \\
\mp \frac{2}{\sqrt{3}} \\
\end{pmatrix}$ & Attractor/Repellor \\
 & $Y = 0$, $Z = \frac{2}{3}$ &$E$ &
 $\begin{pmatrix}
1\\
-1
\end{pmatrix}$ & Saddle \\
& $Y = \pm 1$, $Z=0$ & $dS_{2\pm}$ & N/A & Saddle \\
& $Y = \pm 1$, $Z=1$ & $S_\pm$ & N/A & Repellor/Attractor \\
\hline
 $x = 1$ & $Y = \pm \frac{1}{2}$, $Z = 0$, $R = 0$ & $dS_{1\pm}$ & $\begin{pmatrix}
\mp \frac{4}{\sqrt{3}}\\
\mp \sqrt{3}\\
\mp \frac{2}{\sqrt{3}} \\
\end{pmatrix}$ & Attractor/Repellor \\
 & $Y = 0$, $R = \frac{-2 + 3Z}{-4 + 5Z}$ & $E$ &
 $\begin{pmatrix}
0 \\
\sqrt{\frac{8 - 9Z}{6(1-Z)}}\\
-\sqrt{\frac{8 - 9Z}{6(1-Z)}}
\end{pmatrix}$ & Saddle \\
& $Y = \pm 1$, $Z = 0$, $R = 0$ & $dS_{2\pm}$ & N/A & Saddle \\
& $Y = \pm 1$, $Z = 1$, $R = 1$ & $S_\pm$ & N/A & Repellor/Attractor \\
\hline
 $x = \scriptR$, $Z = 0$ & $Y = \pm \sqrt{\frac{\scriptR}{3 + \scriptR}}$, $R = 0$ & $dS_{1\pm}$ & 
 $\begin{pmatrix}
\mp \frac{4\sqrt{\scriptR}}{\sqrt{3}}\\
\mp \frac{2\sqrt{\scriptR}}{\sqrt{3}} \\
\end{pmatrix} $ & Attractor/Repellor \\
 & $Y = 0$, $R = \frac{\scriptR}{1 + \scriptR}$ & $E$ &
 $\begin{pmatrix}
\frac{2\sqrt{\scriptR}}{3} \\
- \frac{2\sqrt{\scriptR}}{3} 
\end{pmatrix}$ & Saddle \\
& $Y = \pm 1$, $R = 0$ & $dS_{2\pm}$ & N/A & Saddle \\
& $Y = \pm 1$, $R = 1$ & $S_\pm$ & N/A & Repellor/Attractor \\
\hline
 $x = \scriptR$, $R = 0$ & $Y = \pm \sqrt{\frac{\scriptR}{3 + \scriptR}}$, $Z = 0$ & $dS_{1\pm}$ & $\begin{pmatrix}
\mp \sqrt{3\scriptR}\\
\mp \frac{2\sqrt{\scriptR}}{\sqrt{3}}
\end{pmatrix} $ & Attractor/Repellor \\
 & $Y = 0$, $Z = \frac{2\scriptR}{1 + 2\scriptR}$ & $E$ &
 $\begin{pmatrix}
\sqrt{\scriptR} \\
- \sqrt{\scriptR}
\end{pmatrix}$ & Saddle \\
& $Y = \pm 1$, $Z = 0$ & $dS_{2\pm}$ & N/A & Saddle \\
& $Y = \pm 1$, $Z = 1$ & $S_\pm$ & N/A & Repellor/Attractor \\
\hline
 $x = \scriptR$ & $Y = \pm \sqrt{\frac{\scriptR}{3 + \scriptR}}$, $Z = 0$, $R = 0$ & $dS_{1\pm}$ & $\begin{pmatrix}
 \mp \frac{4\sqrt{\scriptR}}{\sqrt{3}} \\
 \mp \sqrt{3\scriptR} \\
\mp \frac{2 \sqrt{\scriptR}}{\sqrt{3}}
\end{pmatrix}$ & Attractor/Repellor \\
 & $Y = 0$, $R = \frac{Z - 2\scriptR + 2Z\scriptR}{-2 + 3Z - 2\scriptR + 2Z\scriptR}$ & $E$ & $\begin{pmatrix}
0 \\
\sqrt{\frac{8\scriptR - Z(1 + 8\scriptR)}{6(1-Z)}} \\
-\sqrt{\frac{8\scriptR - Z(1 + 8\scriptR)}{6(1-Z)}}
\end{pmatrix}$ & Saddle \\
& $Y = \pm 1$, $Z=0$, $R = 0$ & $dS_{2\pm}$ & N/A & Saddle \\
& $Y = \pm 1$, $Z=1$, $R = 1$ & $S_\pm$ & N/A & Repellor/Attractor \\
\hline
 $Z = 0$, $R = 0$ & $Y = \pm\sqrt{\frac{\scriptR}{3 + \scriptR}}$, $x = \scriptR$ & $dS_{1\pm}$ & $\begin{pmatrix}
\mp \frac{2\sqrt{\scriptR}}{\sqrt{3}} \\
\mp \sqrt{3\scriptR}(1-\scriptR)
\end{pmatrix}$ & Attractor/Repellor \\
 & $Y = \pm \frac{1}{2}$, $ x = 1$ & $dS_{2\pm}$ & $\begin{pmatrix}
\mp \frac{2}{\sqrt{3}} \\
\pm \sqrt{3}(1-\scriptR)
\end{pmatrix}$ & Saddle \\
 & $Y = 0$, $x = \frac{1}{6}(1 + 3\scriptR \pm \sqrt{1 - 30\scriptR + 9\scriptR^2}) $ & $E$ &$\begin{pmatrix}
\sqrt{\frac{-1 + 30\scriptR - 9\scriptR^2 \mp (1 + 3\scriptR)\sqrt{1 - 30\scriptR + 9\scriptR^2}}{18}} \\
-\sqrt{\frac{-1 + 30\scriptR - 9\scriptR^2 \mp (1 + 3\scriptR)\sqrt{1 - 30\scriptR + 9\scriptR^2}}{18}}
\end{pmatrix}$  & Saddle, Cusp or Centre \\
& $Y = \pm 1$, $x = \scriptR$ & $dS_{3\pm}$ & N/A & Saddle \\
& $Y = \pm 1$, $x = 1$ & $dS_{4\pm}$ & N/A & Repellor/Attractor \\
\hline
 $Z = 0$ &  $Y = \pm \sqrt{\frac{\scriptR}{3 + \scriptR}}$, $x=\scriptR$, $R = 0$ & $dS_{1\pm}$ & $\begin{pmatrix}
\mp \frac{4\sqrt{R}}{\sqrt{3}} \\
\mp \frac{2\sqrt{\scriptR}}{\sqrt{3}} \\
\mp \sqrt{3\scriptR}(1-\scriptR)
\end{pmatrix}$ & Attractor/Repellor \\
& $Y = 0$, $R = \frac{-x + 3x^2 + 3\scriptR -3x\scriptR}{2 - x + 3x^2 + 3\scriptR - 3x\scriptR}$ & $E$ &
 $\begin{pmatrix}
0 \\
\sqrt{\frac{x(9\scriptR^2 + 18\scriptR - 1) - x^2(27\scriptR + 9) + 18x^3 + 9\scriptR(1-\scriptR)}{6}}\\
-\sqrt{\frac{x(9\scriptR^2 + 18\scriptR - 1) - x^2(27\scriptR + 9) + 18x^3 + 9\scriptR(1-\scriptR)}{6}}
\end{pmatrix}$ & Saddle \\
& $Y = \pm 1$, $x = \scriptR$, $R = 0$ & $dS_{2\pm}$ & N/A & Saddle \\
& $Y = \pm 1$, $x = 1$, $R = 1$ & $S_\pm$ & N/A & Repellor/Attractor \\
\hline
 $R = 0$ &  $Y = \pm \sqrt{\frac{\scriptR}{3 + \scriptR}}$, $x=\scriptR$, $Z = 0$ & $dS_{1\pm}$ & $\begin{pmatrix}
\mp \sqrt{3R} \\
\mp \frac{2\sqrt{\scriptR}}{\sqrt{3}} \\
\mp \sqrt{3\scriptR}(1-\scriptR)
\end{pmatrix}$ & Attractor/Repellor \\
& $Y = 0$, $Z = \frac{-x + 3x^2 + 3\scriptR - 3x\scriptR}{1 - x + 3x^2 + 3\scriptR - 3x\scriptR}$ & $E$ &
 $\begin{pmatrix}
0 \\
\sqrt{\frac{x(7\scriptR + 3\scriptR^2) - x^2(9\scriptR + 4) + 6x^3 + \scriptR(2 - 3\scriptR)}{2}} \\
-\sqrt{\frac{x(7\scriptR + 3\scriptR^2) - x^2(9\scriptR + 4) + 6x^3 + \scriptR(2 - 3\scriptR)}{2}}
\end{pmatrix}$ & Saddle \\
& $Y = \pm 1$, $x = \scriptR$, $Z = 0$ & $dS_{2\pm}$ & N/A & Saddle \\
& $Y = \pm 1$, $x = 1$, $Z = 1$ & $S_\pm$ & N/A & Repellor/Attractor \\
\hline
 $k = 0$ & $Y = \pm \sqrt{\frac{\scriptR}{3 + \scriptR}}$, $x = \scriptR$, $R = 0$ ($Z = 0$) & $dS_{\pm}$ & $\begin{pmatrix}
 \mp \frac{4\sqrt{\scriptR}}{\sqrt{3}}\\
 \mp \sqrt{3\scriptR}\\
 \mp \sqrt{3\scriptR}(1-\scriptR)
\end{pmatrix}$ & Attractor/Repellor\\
 & $Y = 0$, $R = \frac{3x^2 - 3\scriptR x + 3\scriptR}{1 + 3x^2 - 3\scriptR x + 3\scriptR}$
& $E$ & $\begin{pmatrix}
 0 \\
 \sqrt{\frac{x(5\scriptR + 3\scriptR^2) - x^2(2 + 9\scriptR) + 6x^3 + \scriptR(4 - 3\scriptR)}{2}}\\
 -\sqrt{\frac{x(5\scriptR + 3\scriptR^2) - x^2(2 + 9\scriptR) + 6x^3 + \scriptR(4 - 3\scriptR)}{2}}\\
\end{pmatrix}$ & Saddle \\
& $Y = \pm 1$, $x = \scriptR$, $R = 0$ ($Z = 1$) & $S_{1\pm}$ & N/A & Saddle \\
&$Y = \pm 1$, $x = 1$, $R = 1$ ($Z = 0$) & $S_{2\pm}$ & N/A & Repellor/Attractor\\
 \hline

\caption{\label{tab:Sub-manifolds} The sub-manifolds of the system with their fixed points, eigenvalues and stability character. $E$ denotes an Einstein universe, $dS_{\pm}$ an expanding (+) or contracting (-) de-Sitter universe, and $S_{\pm}$ a singularity with infinite expansion (+) or contraction (-). The fixed points at $Y=\pm 1$ representing infinities do not admit a linearization, and therefore eigenvalues cannot be found (denoted by N/A in the Eigenvalues column). We provide details of how we find their stability character in the text.}
 \end{longtable}

 \endgroup

\twocolumngrid

\noindent Evaluating this Jacobian matrix at each of the fixed points and finding the eigenvalues then tells us the stability of the fixed points. If the eigenvalues have non-zero real parts, the fixed point is said to be hyperbolic. If all real parts of the eigenvalues are positive, then the fixed point is a repellor, and if the eigenvalues have negative real parts, then the fixed point is an attractor. If there are positive and negative real parts of the eigenvalues, then the fixed point is a saddle point or a cusp. Finally, if the eigenvalues are purely imaginary, and the real parts are therefore zero, then the fixed point is a centre if the system is linear. For a non-linear system this requires verification from numerical integration and plots of the phase space. The case of complex eigenvalues is not relevant here.

We also classify the fixed points by the type of universe model they represent. The de Sitter models correspond to a cosmological constant and constant energy density of matter and radiation, which in our system corresponds to $x' = Z' = R' = 0$. $Y$ can vary for positively and negatively curved de Sitter models. For flat models, $Y$ is constant, $Y'=0$, giving rise to de Sitter fixed points. These occur at the effective cosmological constants $x = \scriptR$ and $x = 1$, and when $Z = R = 0$. An Einstein universe is static, and in our system is represented by fixed points at $Y' = Y = 0$. Finally, our system admits singularities, which occur when the energy densities of matter and radiation become infinite at $Z = R = 1$.

In the following subsections we consider the sub-manifolds of the system, which simplifies the dynamics by reducing the dimensions of the system \cite{Boehmer2014}. This helps us to understand the conditions for the existence of each fixed point, and understand their nature. A summary of the fixed points of each sub-manifold, along with their eigenvalues and stability character, can be found in Table \ref{tab:Sub-manifolds}. The fixed points labelled $E$ represent a static Einstein universe and the fixed points labelled $dS_{\pm}$ represent spatially flat expanding (+) and contracting (-) de-Sitter models. $S_{\pm}$ denotes singularities with infinite expansion (+) or contraction (-) and infinite energy density. Note we cannot compute the eigenvalues of any fixed point at $Y = \pm 1$, as the Jacobian becomes singular and we therefore cannot Taylor expand around them. However, from the Raychaudhuri equation \eqref{CompactHubbleRateDE}, we find that $Y'$ is negative along the de Sitter lines $x = \scriptR$ and $x = 1$ around the fixed points at $Y = \pm 1$.

\subsection{$\Lambda$CDM Dynamics}

We begin with the sub-manifolds with an effective cosmological constant and so are analogous to $\Lambda$CDM, except here $x = 1$ and $x = \scriptR$ are asymptotic values rather than true constants. In each of these cases, $x' = 0$ \eqref{CompactDarkEnergyDE}.

\subsubsection{$x = 1$}

We begin with the $x = 1$ sub-manifold. The resulting dynamics is 3-dimensional (3-D), given by the equations for dark matter \eqref{CompactDarkMatterDE}, radiation \eqref{CompactRadiationDE}, and the Hubble variable \eqref{CompactHubbleRateDE}, which reduces to  

\begin{multline}
    Y' = -Y^2\sqrt{1-Y^2} - \frac{(1-Y^2)^\frac{3}{2}}{6} \\ \times
    \left(\frac{Z}{1-Z} + \frac{2R}{1-R} - 2\right)\,.
    \label{eqn:Y'x1}
\end{multline}

\noindent In this case, the Friedmann equation \eqref{eqn:Friedmann_Dimensionless_Compact} becomes

\begin{equation}
    \frac{Y^2}{1-Y^2}=\frac{1}{3}+\frac{Z}{3(1-Z)}+\frac{R}{3(1-R)}-\frac{k}{\rho_* a^2}\,,
    \label{eqn:Friedmann_x=1}
\end{equation}

\noindent and the potential \eqref{eqn:compact_U} is

\begin{equation}
    U = -\frac{a^2}{6}\left(1 + \frac{Z}{1-Z} + \frac{R}{1-R}\right)\,.
    \label{eqn:U_x=1}
\end{equation}

\noindent For the sub-manifolds where $x$ is a constant, we express $a$ in terms of $Z$. Integrating $a'$ \eqref{eqn:a'} with respect to $Z$ \eqref{CompactDarkMatterDE}, we find

\begin{equation}
    a = \left(\frac{1-Z}{Z}\frac{Z_0}{1-Z_0}\right)^\frac{1}{3} \,.
    \label{eqn:a(Z)}
\end{equation}

\noindent We can then solve for the fixed points of the system. There are de Sitter fixed points where $Z = R = 0$, and $Y = \pm 1/2$ and $Y = \pm 1$. The fixed points at $Y = \pm 1$ and at $Z = R = 0$ are coordinate singularities of the de Sitter spacetime when represented as an FLRW, see \cite{Hawking1973}. At $Z = R = 1$, \eqref{CompactDarkMatterDE} and \eqref{CompactRadiationDE} have fixed points, and at $Y = \pm 1$ \eqref{eqn:Y'x1} has fixed points. Together, the asymptotic points at $Z = R = 1$ and $Y = \pm 1$ represent singularities, with infinite expansion (+) or contraction (-) and infinite energy density. At the Einstein fixed point, $Y=0$ and the condition $Y' = 0$ reduces Eq. \eqref{eqn:Y'x1} to a constraint between $Z$ and $R$:

\begin{equation}
     \frac{Z_E}{1-Z_E} + \frac{2R_E}{1-R_E} - 2 = 0\,,
     \label{eqn:x1_Einstein}
\end{equation}

\noindent where the subscript $E$ refers to the values of variables at the Einstein point. On the other hand, matter and radiation do not evolve independently, and from \eqref{rhomdot} and \eqref{rhordot}, they can be related by 

\begin{equation}
    \rho_r = \frac{1}{(3H_0^2)^{1/3}}\frac{\Omega_r}{\Omega_m^{4/3}} \rho_m^{4/3},
\end{equation}

\noindent where $H_0$, $\Omega_r$ and $\Omega_m$ can be fixed using Planck values \cite{Planck2018VI}. This results in \eqref{CompactDarkMatterDE} and \eqref{CompactRadiationDE} admitting a first integral $c_{rz}$, which can be used to write a relation between $R$ and $Z$

\begin{equation}
    \frac{R}{1-R} = c_{rz} \left(\frac{Z}{1-Z}\right)^\frac{4}{3} ,
    \label{eqn:R(Z)}
\end{equation}

\noindent where $c_{rz}$ depends on $\Omega_m$, $\Omega_r$ and $\rho_*$. Together, \eqref{eqn:x1_Einstein} and \eqref{eqn:R(Z)} give the values of $Z$ and $R$ at the Einstein fixed point for a given value of $c_{rz}$. Once the first integral $c_{rz}$ is fixed, a surface in phase space is defined and the dynamics is effectively 2-dimensional (2-D). In order to fix $c_{rz}$, we first have to give a value to $\rho_* = \rho_\Lambda / \scriptR$, where the future asymptotic value $\rho_\Lambda$ is by definition some fraction of our current observed dark energy density,

\begin{equation}
    \rho_\Lambda = \alpha \rho_{x,0}\,,
    \label{eqn:rho_Lambda}
\end{equation}

\noindent and $0 < \alpha < 1$. For the purpose of our qualitative analysis, we fix $\alpha = 0.5$ as currently we are not far from this asymptotic value as our Universe is already accelerating. For a reasonable model to solve the cosmological constant problem, we would fix $10^{-120} < \scriptR < 10^{-60}$ so the dark energy evolved between the high estimate of the contributions to the effective cosmological constant from QFT and the low observed value. However, the dynamics cannot be depicted in plots of the phase space for these small values of $\scriptR$. We therefore choose $\scriptR = 0.05$ in order to illustrate the evolution of our variables, given that the value of $\scriptR$ will not qualitatively change the dynamics. We therefore find

\begin{multline}
    c_{rz} = \frac{\Omega_r}{(\Omega_m)^\frac{4}{3}}\left(\frac{\alpha \Omega_\Lambda}{\scriptR}\right)^\frac{1}{3} 
    = \Omega_r\left(\frac{\Omega_\Lambda}{\Omega_m^4}\right)^\frac{1}{3} \left(\frac{\alpha}{\scriptR}\right)^\frac{1}{3} 
    \\
    \simeq 0.00083 \,,
\end{multline}

\noindent where we have used $\Omega_\Lambda = 0.6889$ and $\Omega_m = 0.3111$, and calculated $\Omega_r = 9.1824 \times 10^{-5}$ using $\Omega_m$ and the redshift of matter-radiation equality $z = 3387$ \cite{Planck2018VI}.  

\begin{figure}
\begin{center}
\includegraphics[width=90mm]{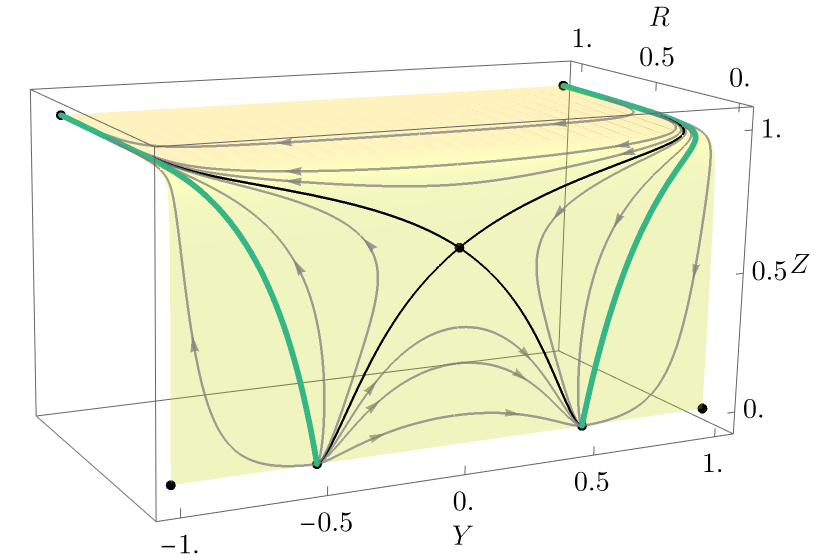}
\end{center}
\caption{The phase space for the $x = 1$ sub-manifold, with the dynamics shown on the yellow $c_{rz}$ surface, given by \eqref{eqn:R(Z)}. The green curves represent flat models asymptotic to the de Sitter fixed points at $Z = 0$ and singularities at $Z = 1$. The fixed point at $Y=0$ (a saddle) represents the Einstein model, and the black separatrices asymptotic to it separate the different types of closed models in the phase space.}
\label{fig:x1SubMan}
\end{figure}

The phase space for the $x = 1$ sub-manifold is shown in Fig. \ref{fig:x1SubMan}, where the dynamics takes place on the first integral surface $c_{rz}$, shown in yellow. The green outer most thick curve is the flat Friedmann separatrix (FFS), which separates the closed models (in between the green curves) from the open models. This separatrix on the 2-D surface is a sub-set of the general separatrix hypersurface, which has the general form

\begin{equation}
    \frac{Y^2}{1-Y^2} = \frac{x}{3} + \frac{Z}{3(1-Z)} + \frac{R}{3(1-R)},
    \label{eqn:FFS}
\end{equation}

\noindent where here $x = 1$ and $R$ and $Z$ are related by \eqref{eqn:R(Z)}. The inner most curve is the closed Friedmann separatrix (CFS), which in general is the hypersurface consisting of the Einstein fixed points present in the phase space, and trajectories asymptotic to them. This shows the boundaries between the different types of closed model. The CFS has the general form

\begin{multline}
    \frac{Y^2}{1-Y^2} = \frac{x}{3} + \frac{Z}{3(1-Z)} + \frac{R}{3(1-R)} \\ - \frac{1}{a^2}\left[\frac{x_E}{3} + \frac{Z_E}{3(1-Z_E)} + \frac{R_E}{3(1-R_E)}\right]\,,
    \label{eqn:CFS}
\end{multline}

\noindent where again here $x =x_E= 1$ and the relations in Eq. \eqref{eqn:a(Z)} and Eq. \eqref{eqn:R(Z)} apply.

\begin{figure}
\begin{center}
\includegraphics[width=80mm]{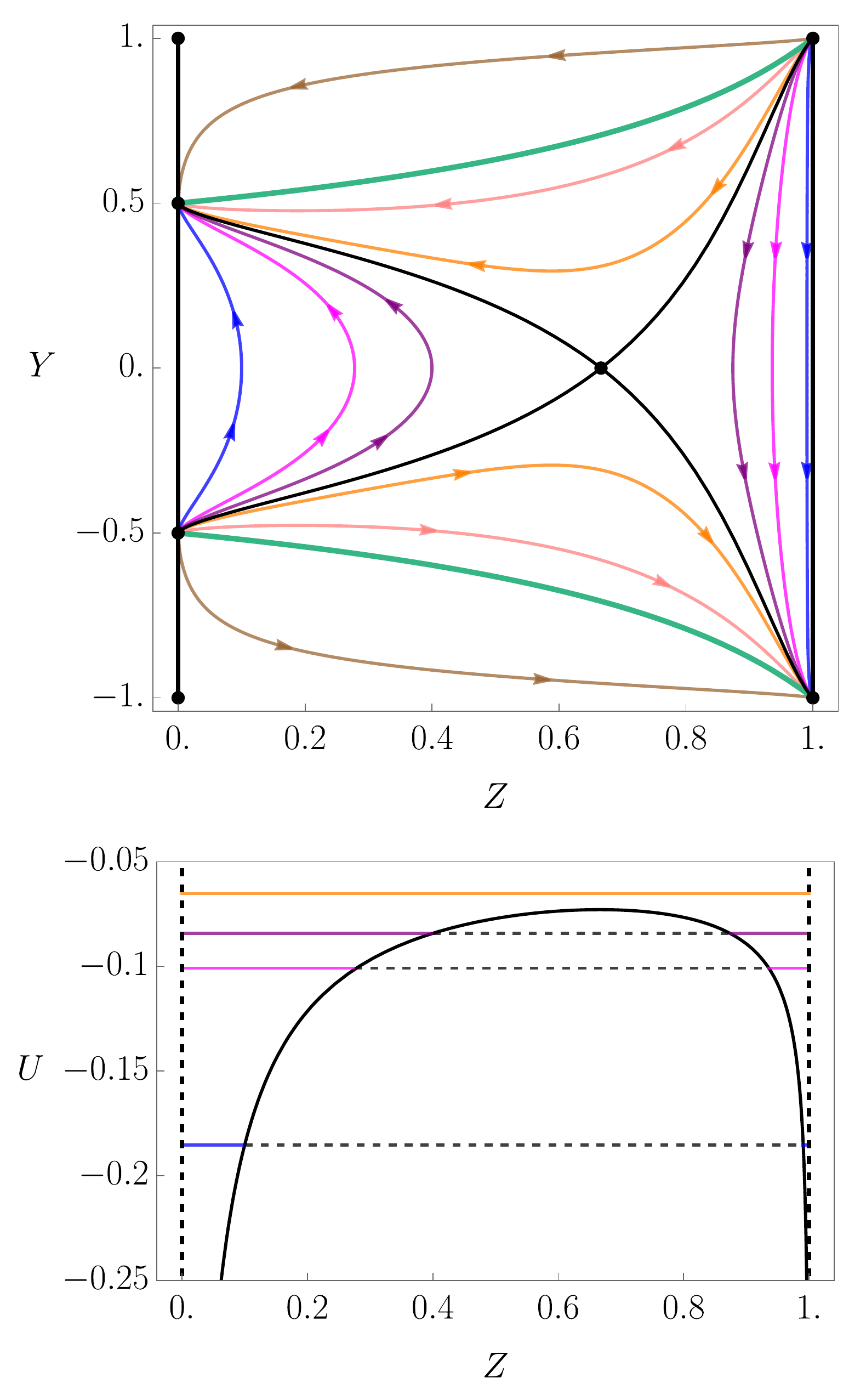}
\end{center}
\caption{$x = 1$ sub-manifold, top panel:  the projection on the $Z$-$Y$ plane of the 2-D dynamics taking place on the first integral surface in the 3-D phase space in Fig. \ref{fig:x1SubMan}; the qualitative behaviour is typical of any $\Lambda$CDM model, independently of the parameter values. Bottom panel: the corresponding potential in Eq. \eqref{eqn:U_x=1}, where trajectories of the same colour in the two panels correspond to each other. Bounded trajectories are below the maximum of the potential, and unbounded are above.}
\label{fig:x1SubManProjection}
\end{figure}

The 2-D dynamics taking place on the first integral surface in the 3-D phase space can also be projected onto the $Z$-$Y$ plane in order to make the dynamics more visible, again using \eqref{eqn:R(Z)}.
The projected sub-manifold can be seen in Fig. \ref{fig:x1SubManProjection}. The vertical black lines along $Z = 0$ and $Z = 1$ show the de Sitter models in the phase space, where $x' = Z' = R' = 0$. Expanding (contracting) open and flat models evolve from (to) a singularity to (from) a flat de Sitter fixed point. Closed models outside the CFS evolve in the same way. Within the CFS there are bouncing models, which contract until they reach a maximum energy density at a minimum $a$ where they bounce, and then expand again. The bounce occurs when the dark energy component is dominant over the matter and radiation, and the magnitude of the curvature term is equal to the sum of all other contributions to the Friedmann equation \eqref{eqn:Friedmann_x=1}, giving $Y = 0$. There are also turn-around models, which expand until they reach a minimum energy density at a maximum $a$ before contracting again, that evolve between the two singularities. The turn-around occurs at $Y = 0$, when the dark matter and radiation are dominant over the dark energy, and the magnitude of the curvature term becomes large enough to equal the sum of all other contributions to the Friedmann equation \eqref{eqn:Friedmann_x=1}. Note that for all trajectories in the $Z$-$Y$ plane, the expansion (contraction) is accelerating ($a'' > 0$) to the left of the Einstein point. Therefore in a $\Lambda$CDM cosmology, such as the one we are illustrating here, bouncing models are always accelerating, and re-collapsing models are never accelerating.

There are also two other 2-D submanifolds at $x = 1$; one for $Z = 0$ (radiation only) and the other when $R = 0$ (matter only). The dynamics of these two sub-cases is qualitatively the same as in Fig. \ref{fig:x1SubManProjection}. In the $Z = 0$ case the Einstein point occurs at $R_E = 1/2$, and in the $R = 0$ case it occurs at $Z_E = 2/3$. 

\subsubsection{$x = \scriptR$}

Next we consider the case where $x = \scriptR$. The remaining 3-dimensional dynamics is given by the equations for the dark matter \eqref{CompactDarkMatterDE}, radiation \eqref{CompactRadiationDE}, and the Hubble function \eqref{CompactHubbleRateDE}, which reduces to

\begin{multline}
    Y' = -Y^2\sqrt{1-Y^2} - \frac{(1-Y^2)^\frac{3}{2}}{6} \\
    \times \left(\frac{Z}{1-Z} + \frac{2R}{1-R} - 2\scriptR\right)\,.
    \label{eqn:Y'xR}
\end{multline}

\noindent The Friedmann equation \eqref{eqn:Friedmann_Dimensionless_Compact} when $x = \scriptR$ becomes

\begin{equation}
    \frac{Y^2}{1-Y^2}=\frac{\scriptR}{3}+\frac{Z}{3(1-Z)}+\frac{R}{3(1-R)}-\frac{k}{\rho_* a^2}\,,
    \label{eqn:Friedmann_x=scriptR}
\end{equation}

\noindent and the potential \eqref{eqn:compact_U} is

\begin{equation}
    U = -\frac{a^2}{6}\left(\scriptR + \frac{Z}{1-Z} + \frac{R}{1-R}\right)\,.
    \label{eqn:U_x=scriptR}
\end{equation}

\noindent The dynamics of this system is qualitatively the same as in Fig. \ref{fig:x1SubMan}. The singularities are present at $Z = R = 1$ and $Y = \pm 1$, and there are de Sitter points at $Y = \sqrt{\scriptR/(3 + \scriptR)}$ and $Y = \pm 1$ along $Z = R = 0$. At the Einstein fixed point, the Hubble expansion variable \eqref{eqn:Y'xR} reduces to

\begin{equation}
     \frac{Z_E}{1-Z_E} + \frac{2R_E}{1-R_E} - 2\scriptR = 0\,,
\end{equation}

\noindent which is solved numerically using \eqref{eqn:R(Z)} to find the values of $Z_E$ and $R_E$ at this point.

As before, this sub-manifold can be reduced to 2-D at $Z = 0$ and at $R = 0$, which are both qualitatively the same as in Fig. \ref{fig:x1SubManProjection}. The de Sitter points and singularities are as outlined above. The Einstein point in the $Z = 0$ case occurs at $R_E = \scriptR/(1 + \scriptR)$, and in the $R = 0$ case at $Z_E = 2\scriptR/(1 + 2\scriptR)$.

\subsection{Dynamic Dark Energy}

We now consider the sub-manifolds with dynamic dark energy so that $x$ is not fixed, and therefore $x' \neq 0$ in Eq. \eqref{CompactDarkEnergyDE}. 

\subsubsection{$Z = 0$ and $R = 0$}

We consider the sub-manifold where matter and radiation are not present, and there is only dark energy. In this case $Z' = 0$ in Eq. \eqref{CompactDarkMatterDE} and $R' = 0$ in Eq. \eqref{CompactRadiationDE}. The remaining dynamics is 2-D and given by the equations for the dark energy \eqref{CompactDarkEnergyDE} and the Hubble expansion variable \eqref{CompactHubbleRateDE} which becomes

\begin{multline}
    Y' = -Y^2\sqrt{1-Y^2} - \frac{(1-Y^2)^\frac{3}{2}}{6} \\
    \times [x(1+3\scriptR) - 3\scriptR - 3x^2].
    \label{xYY'}
\end{multline}

\noindent When $Z = R = 0$, the Friedmann equation \eqref{eqn:Friedmann_Dimensionless_Compact} becomes

\begin{equation}
    \frac{Y^2}{1-Y^2}=\frac{x}{3} - \frac{k}{\rho_* a^2}\,,
    \label{eqn:Friedmann_Z=R=0}
\end{equation}

\noindent and the potential \eqref{eqn:compact_U} is

\begin{equation}
    U = -\frac{xa^2}{6}\,.
    \label{eqn:U_Z=R=0}
\end{equation}

\noindent In order to plot the potential and the CFS in the phase space, we must express the scale factor $a$ in terms of $x$. Integrating $a'$ \eqref{eqn:a'} with respect to $x$ \eqref{darkenergyDE}, we find

\begin{equation}
    a(x) = \left(c_a \frac{x-\scriptR}{1-x}\right)^{\frac{-1}{3(1-\scriptR)}}\,,
    \label{eqn:a(x)}
\end{equation}

\noindent i.e. the inverse of Eq. \eqref{eqn:x(a)}, where $c_a$ is as in Eq. \eqref{eqn:ca}. There are de Sitter points at $Y = \sqrt{\scriptR/(3 + \scriptR)}$ and $Y = \pm 1$ along $x = \scriptR$, and at $Y = \pm 1/2$ and $Y = \pm 1$ along $x = 1$. The fixed points at $Y = \pm 1$ and at $x = \scriptR$ and $x = 1$ are coordinate singularities of the de Sitter spacetime when represented as an FLRW, see \cite{Hawking1973}. At the Einstein fixed point, $Y = 0$, Eq. \eqref{xYY'} reduces to

\begin{equation}
    x_E(1+3\scriptR) - 3\scriptR - 3x_E^2 = 0.
\end{equation}

\noindent Rearranging for $x_E$, we find

\begin{equation}
    x_{E\pm} = \frac{1+3\scriptR}{6} \pm \sqrt{\frac{(1+3\scriptR)^2}{36}-\scriptR}.
    \label{xYx}
\end{equation}

\noindent Given the condition $0 < \scriptR < 1$, both roots in Eq. \eqref{xYx} will be positive, provided that the term under the square root is greater or equal to zero. Thus, for Einstein points to exist, we find

\begin{equation}
    \scriptR \leq  \scriptR_M \equiv   \frac{5}{3} - \frac{2\sqrt{6}}{3}\,.
\end{equation}

\noindent Therefore, there are three cases for this sub-manifold: no Einstein points are admitted when $\scriptR >\scriptR_M$, two Einstein points exist when $\scriptR < \scriptR_M$, and there is a limiting case in between when $\scriptR = \scriptR_M$ where one Einstein point is admitted. The open and flat models evolve in the same way
for each case. Expanding (contracting) open models evolve between two de Sitter fixed points, from (to) an
open geometry to (from) a flat geometry. Expanding (contracting) flat models evolve between two de Sitter
fixed points along the FFS. In general, all closed models admit a bounce, however the qualitative behaviour changes depending on the number of Einstein points admitted.

\begin{figure}
\begin{center}
\includegraphics[width=80mm]{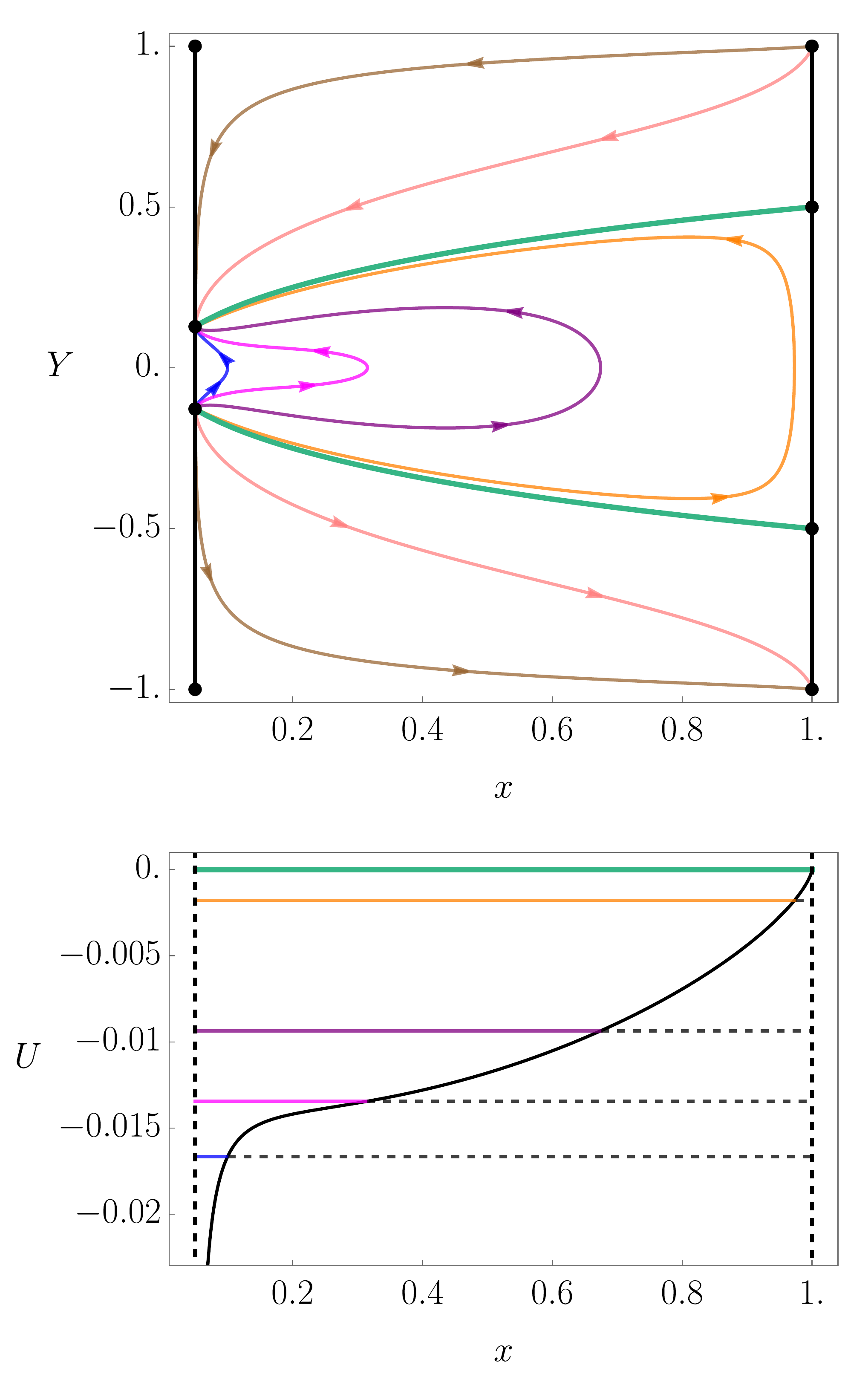}
\end{center}
\caption{Top panel: the phase space of the $Z=0$, $R=0$ sub-manifold, with $\scriptR > \scriptR_M$. Bottom panel: corresponding potential in Eq. \eqref{eqn:U_Z=R=0}, where trajectories of the same colour in the two panels correspond to each other. In this case, there are no Einstein points present, and so all closed models bounce, however there is never a decelerating phase.}
\label{fig:Z0R00E}
\end{figure}

\begin{figure}
\begin{center}
\includegraphics[width=80mm]{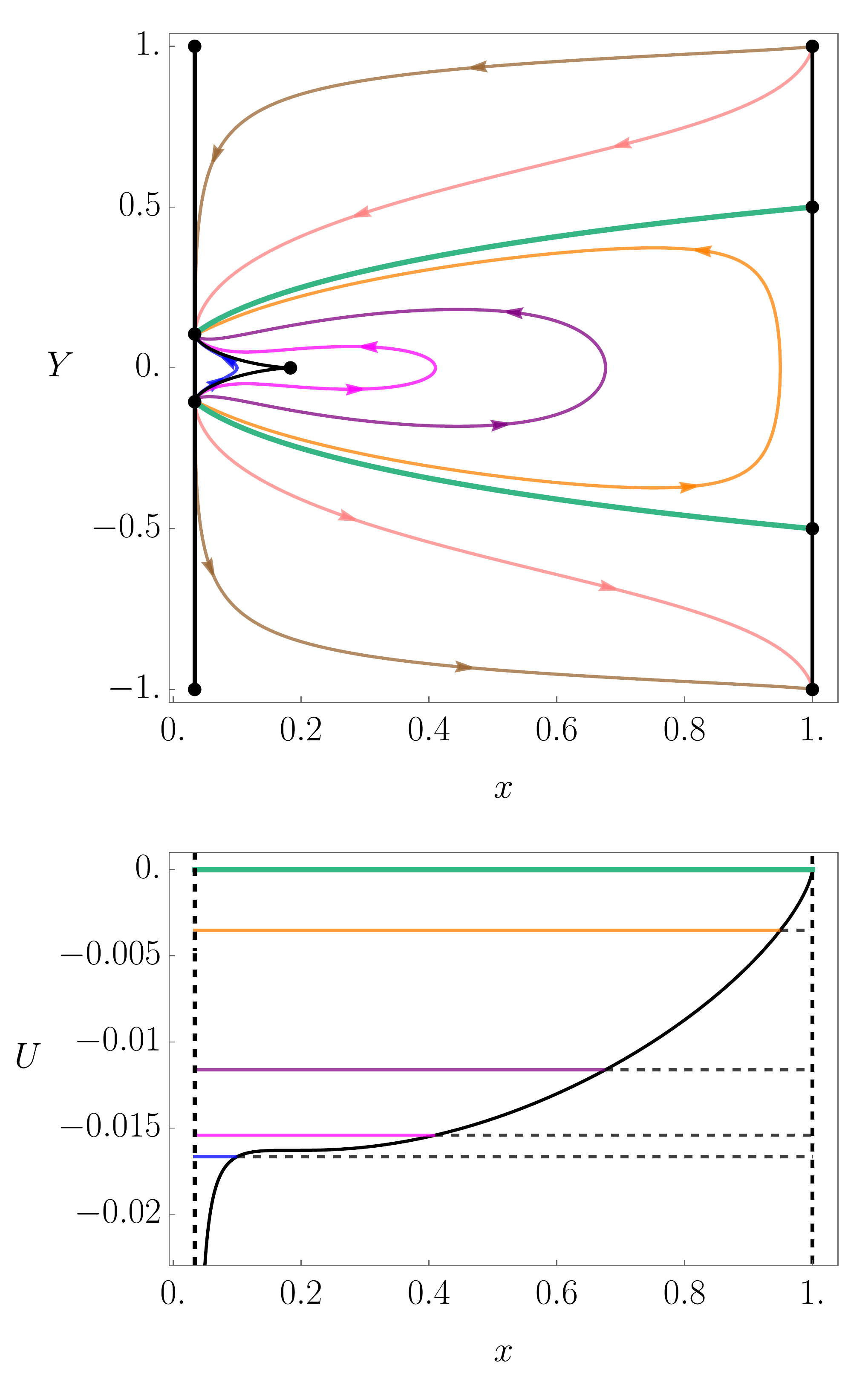}
\end{center}
\caption{Top panel: the phase space of the $Z=0$, $R=0$ sub-manifold, with $\scriptR = \scriptR_M$. Bottom panel: corresponding potential in Eq. \eqref{eqn:U_Z=R=0}, where trajectories of the same colour in the two panels correspond to each other. One Einstein point (cusp) is present, and all closed models bounce, however there is never a decelerating period.}
\label{fig:Z0R01E}
\end{figure}

The $\scriptR > \scriptR_M$ case is shown in Fig. \ref{fig:Z0R00E}, where the system admits no Einstein fixed points. The two vertical black lines along $x = \scriptR$ and $x = 1$ highlight the de Sitter models in the phase space. All closed models bounce, evolving between two de Sitter fixed points. The limiting $\scriptR = \scriptR_M$ case is shown in Fig. \ref{fig:Z0R01E}. Here, the system admits one Einstein point, which is a cusp, so all closed models bounce, evolving between two de Sitter fixed points. 

\begin{figure}
\begin{center}
\includegraphics[width=80mm]{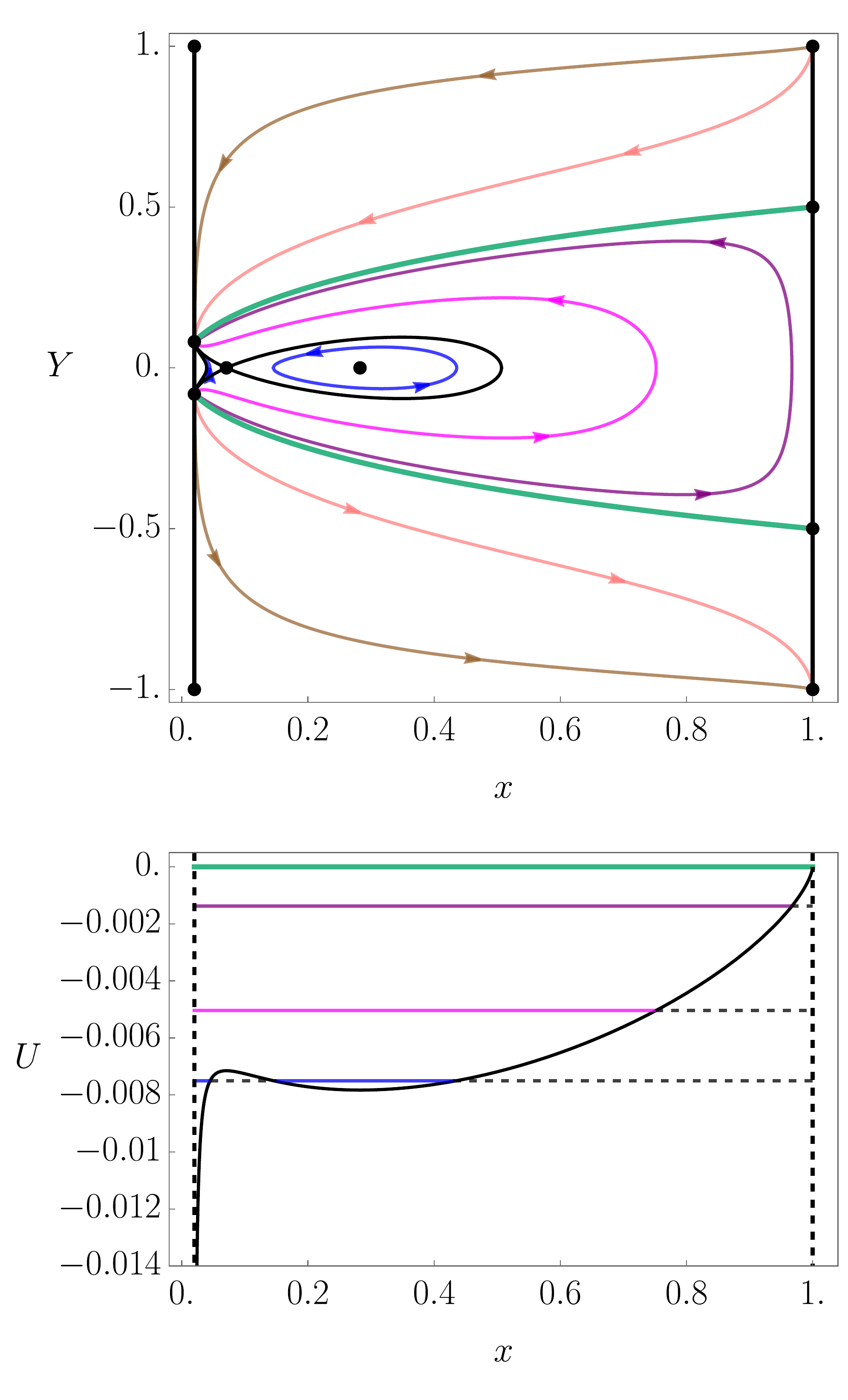}
\end{center}
\caption{Top panel: the phase space of the $Z=0$, $R=0$ sub-manifold, with $\scriptR < \scriptR_M$. Bottom panel: corresponding potential in Eq. \eqref{eqn:U_Z=R=0}, where trajectories of the same colour in the two panels correspond to each other. Two Einstein points at $x_E \simeq 0.07$ (a saddle) and $x_E \simeq 0.28$ (a centre) are present, therefore all closed models bounce, either once or in repeated cycles. Deceleration occurs for $x$ between the two Einstein points. The closed models outside the CFS evolve with an initial and final acceleration, with a decelerating period in between.}
\label{fig:Z0R02E}
\end{figure}

Finally, the case where $\scriptR < \scriptR_M$ is shown in Fig. \ref{fig:Z0R02E}. Two Einstein points are admitted: the fixed point at $x_E \simeq 0.07$ is a saddle point, which is part of the CFS (the black loop in Fig. \ref{fig:Z0R02E}), and the other at $x_E \simeq 0.28$ is a centre. Closed models within the CFS either bounce once, evolving between two de Sitter fixed points, or are cyclic around the centre fixed point, repeatedly contracting until they bounce, then expanding until they turn-around. Closed models outside the CFS also bounce and evolve between two de Sitter fixed points. These closed models outside the CFS evolve with an early- and late-time acceleration, connected by a decelerating period for $x$ between the two Einstein points. This is the only case for this set of sub-manifolds to have a period of deceleration, as here $\scriptR$ is small enough for the effective EoS to be larger than $-1/3$ for $x$ values between the two Einstein points, which violates the condition for acceleration in Eq. \eqref{eqn:accn_condition}.

\subsubsection{$Z = 0$, $R = 0$}

A three-dimensional sub-manifold exists where $Z = 0$, and another when $R = 0$. Adding only dark matter to the system has the same effect as adding only radiation; the difference between the two are the values of the first integral of $Z(x)$ and $R(x)$ needed to obtain each case. The dynamics is qualitatively the same as the cases we present of the full system in section \ref{sec:Full_System}, where we include both dark matter and radiation in our analysis, therefore we do not include these sub-manifolds explicitly here.

\subsubsection{Spatially flat models: $k = 0$}

\begin{figure}
\begin{center}
\includegraphics[width=80mm]{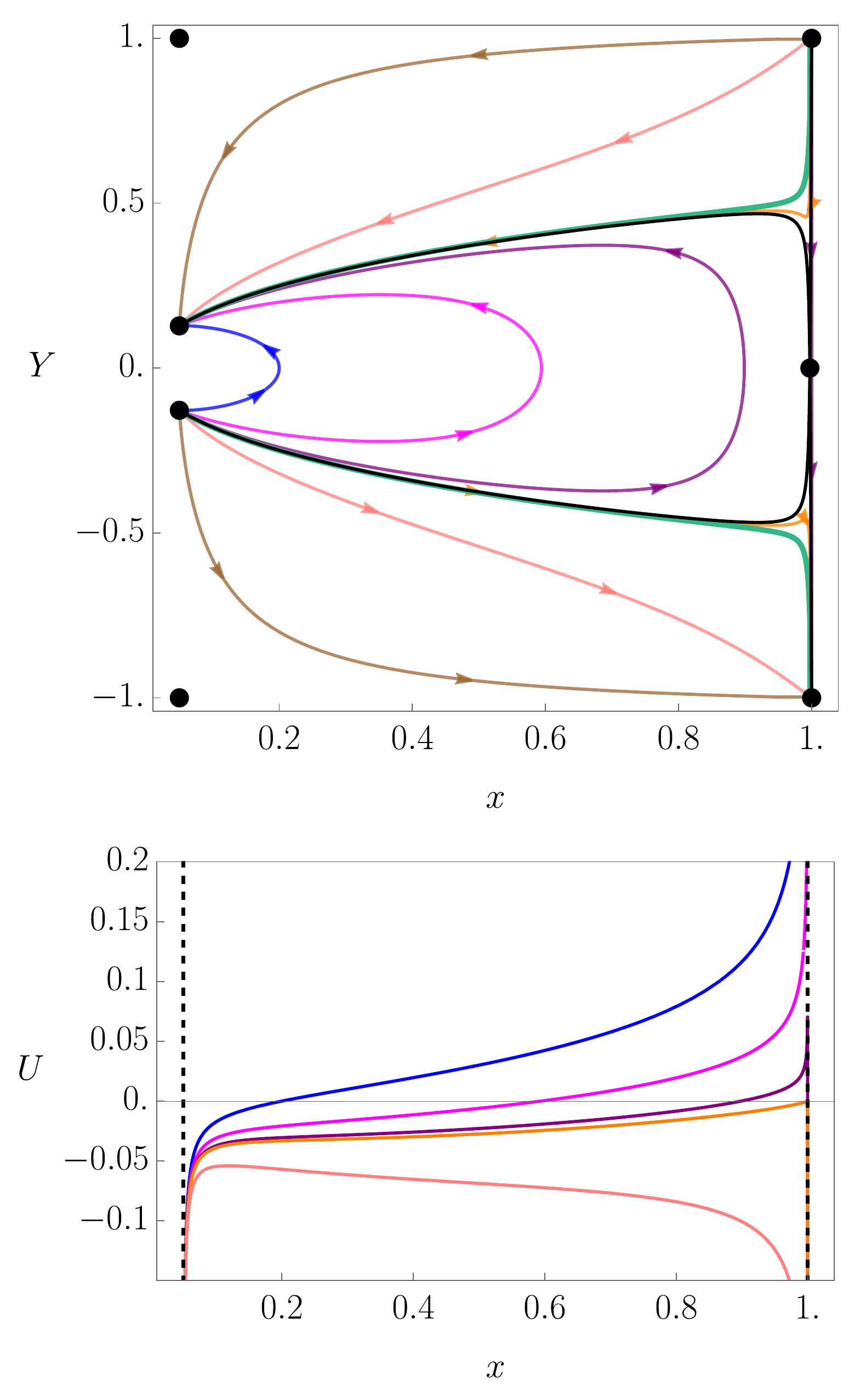}
\end{center}
\caption{Top panel: the phase space of the $k=0$ sub-manifold. Bottom panel: corresponding potential in Eq. \eqref{eqn:compact_U}, where trajectories of the same colour in the two panels correspond to each other. Models within the green $Z = 0$ separatrix always have negative dark matter energy density $Z < 0$. The shape of the potential changes with the initial conditions, and as these are all flat models $k = 0$ the total energy of the system $E = 0$ \eqref{eqn:E}. Therefore, the behaviour of each trajectory can be seen along $U = 0$ for each potential.}
\label{fig:k0}
\end{figure}

Finally, we consider the sub-manifold where the models are all spatially flat, shown in Fig. \ref{fig:k0}, where we have compactified the 4-dimensional (4-D) phase space to 2-D on the $x$-$Y$ plane. In order to do this, we express $z$ and $r$ as functions of $x$. From Eq. \eqref{darkmatterDE} and Eq. \eqref{darkenergyDE}, we find

\begin{equation}
    z(x)=c_z\left(\frac{x-\scriptR}{1-x}\right)^\frac{1}{1-\scriptR}\,,
    \label{eqn:z(x)}
\end{equation}

\noindent where the constant of integration is

\begin{equation}
    c_z=z_0 \left(\frac{1-x_0}{x_0-\scriptR}\right)^\frac{1}{1-\scriptR}\,.
\end{equation}

\noindent Here, to express $c_z$, we set $z_0$ using the Friedmann equation \eqref{eqn:Friedmann_Dimensionless_Compact} when $k=0$:

\begin{equation}
    \frac{Y^2}{1-Y^2}=\frac{x}{3}+\frac{Z}{3(1-Z)}+\frac{R}{3(1-R)}\,.
    \label{eqn:Friedmann_k=0}
\end{equation}

\noindent We can then express $z_0$ as

\begin{equation}
    z_0 = \frac{Z_0}{(1-Z_0)}=\frac{3Y_0^2}{1-Y_0^2} - x_0 - \frac{R_0}{(1-R_0)}\,.
    \label{eqn:Friedmann_z0}
\end{equation}

Then to express $r$ in terms of $x$, from Eq. \eqref{radiationDE} and Eq. \eqref{darkenergyDE} we find

\begin{equation}
    r(x)=c_r\left(\frac{x-\scriptR}{1-x}\right)^\frac{4}{3(1-\scriptR)}\,,
    \label{eqn:r(x)}
\end{equation}

\noindent where the constant of integration is

\begin{equation}
    c_r=r_0 \left(\frac{1-x_0}{x_0-\scriptR}\right)^\frac{4}{3(1-\scriptR)}\,.
\end{equation}

\noindent We set $c_r$ in a similar way to $c_{rz}$, using the density parameters \cite{Planck2018VI}, and our definitions of $\scriptR$ \eqref{eqn:dimensionless_variables} and $\rho_\Lambda$ \eqref{eqn:rho_Lambda}. Keeping our choice of $\scriptR = 0.05$, we then find

\begin{equation}
    c_r = \frac{\scriptR}{\alpha} \frac{\Omega_r}{\Omega_\Lambda}\left(\frac{\alpha-\scriptR}{\scriptR(1-\alpha)}\right)^\frac{4}{3(1 - \scriptR)} \simeq 0.00077\,.
    \label{eqn:c_r}
\end{equation}

The potential is as in Eq. \eqref{eqn:compact_U}. The shape of the potential in this case changes with the initial conditions, and as $k = 0$ the total energy of the system $E = 0$ \eqref{eqn:E}. Therefore the behaviour of each trajectory in the phase space can be seen along $U = 0$ in the bottom panel of Fig. \ref{fig:k0}. Here, the green separatrix represents models where $Z = 0$, so there is no dark matter component, and the innermost black curve is the CFS. There are no de Sitter lines along $x = \scriptR$ or $x = 1$ as in the flat case, $Z$ varies for constant $x$. We do not consider models within the FFS as $Z < 0$ in this region. The models outside this separatrix always have positive dark matter energy density, $Z > 0$. Expanding (contracting) models evolve from (to) an initial (a future) singularity to (from) a de Sitter fixed point. Since these models do not exhibit any bouncing behaviour, they are not interesting with respect to our investigation.

\section{The Full System} \label{sec:Full_System}

To show the full range of dynamics, we project the full 4-D dynamics to 2-D on the $x$-$Y$ plane, by expressing $z$ and $r$ in terms of $x$ as in Eq. \eqref{eqn:z(x)} and Eq. \eqref{eqn:r(x)}, respectively. We set $c_r$ as in Eq. \eqref{eqn:c_r}, keeping $\scriptR = 0.05$ in order for the dynamics to be visible, and to be able to analyse the full range of the dynamics. We do not fix $c_z$ in the full system, but instead investigate how the dynamics varies for different values of this parameter. In order to produce plots of the phase space, we must also express the scale factor $a$ in terms of $x$, using Eq. \eqref{eqn:a(x)}. The Friedmann equation \eqref{eqn:Friedmann_Dimensionless_Compact} can therefore be written just in terms of $x$ and $Y$, and the potential \eqref{eqn:U} can be written in terms of $x$ only. The fixed points for the full system are given in Table \ref{tab:Fixed_Points}. 

As before, $E$ represents a static Einstein universe, $dS_{\pm}$ represents spatially flat expanding (+) and contracting (-) de-Sitter models, and $S_{\pm}$ represents singularities with infinite expansion (+) or contraction (-) and infinite energy density. Note that the $dS_{2\pm}$ fixed points are coordinate singularities of the de Sitter
spacetime when represented as an FLRW, see \cite{Hawking1973}.

Table \ref{tab:Eigenvalues} shows the eigenvalues of the fixed points for this system, and the linear stability classification for each point is given in Table \ref{tab:Linear_Stability}. Again, we cannot include the eigenvalues of the fixed points at $Y = \pm 1$ in Table \ref{tab:Eigenvalues}, as the Jacobian becomes singular. Therefore, we find that the singularity at $Y = +1$ is a repellor and the singularity at $Y = -1$ is an attractor, and the de Sitter points at $Y = \pm 1$ are saddle points, with trajectories moving away from the $Y = +1$ point and towards the $Y = -1$ point along $x = \scriptR$. Note that the linear stability analysis of the Einstein points needs to be done for each individual case, as these can be a centre, a cusp or a saddle point, whereas the linear stability of the de-Sitter points and singularities are general for all cases considered here.

\newpage
\onecolumngrid

\begin{table}[h!]
\centering
 \begin{tabular}{||c | c c c c||} 
 \hline
 Name & $x$ & $Y$ & $Z$ & $R$\\ [0.5ex] 
 \hline\hline
 $E$ & $x$ & $0$ & $Z$ & $\frac{-3x^2(1-Z)+Z(1+3\scriptR)-3\scriptR +x(1-Z)(1+3\scriptR)}{-2-3x^2(1-Z)-3\scriptR+3Z(1+\scriptR)+x(1-Z)(1+3\scriptR)}$\\
 dS$_{1\pm}$ & $\scriptR$ & $\pm\sqrt{\frac{\scriptR}{3 + \scriptR}}$ & $0$ & $0$ \\
 dS$_{2\pm}$ & $\scriptR$ & $\pm 1$ & $0$ & $0$ \\
 S$_{\pm}$ & $1$ & $\pm 1$ & $1$ & $1$ \\
 \hline
 \end{tabular}
 \caption{\label{tab:Fixed_Points} The fixed points of the full system. $E$ denotes an Einstein universe, $dS_{\pm}$ an expanding (+) or contracting (-) de-Sitter universe, and $S_{\pm}$ a singularity with infinite expansion (+) or contraction (-).}
\end{table}

\begin{table}[h!]
\centering
 \begin{tabular}{|| c | c c c c ||} 
 \hline
  Name & $\lambda_1$ & $\lambda_2$ & $\lambda_3$ & $\lambda_4$ \\ 
 \hline\hline
 $E$ &  0 & 0 & $\frac{\sqrt{18 x^3 (Z-1)-9 x^2 (Z-1) (3 \scriptR+1)+x (Z-1) \left(9 \scriptR^2+18 \scriptR-1\right)+Z \left(-9 \scriptR^2+9 \scriptR+1\right)+9 (\scriptR-1) \scriptR}}{\sqrt{6} \sqrt{Z-1}}$ & $-\lambda_3$ \\
 $dS_{1\pm}$ & $\mp 4 \sqrt{\frac{\scriptR}{3}}$ & $\mp \sqrt{3\scriptR}$ & $\mp 2 \sqrt{\frac{\scriptR}{3}}$ & $\mp \sqrt{3\scriptR}(1 - \scriptR)$ \\
 \hline
 \end{tabular}
 \caption{\label{tab:Eigenvalues} The eigenvalues for the fixed points of the full system given in Table \ref{tab:Fixed_Points}.}
\end{table}

\twocolumngrid

\begin{table}[h!]
\centering
 \begin{tabular}{||c | c ||} 
 \hline
  Name & Stability Character \\ 
 \hline\hline
 $E$ & Centre, Saddle or Cusp \\
 $dS_{1+}$ & Attractor \\
 $dS_{1-}$ & Repellor \\
 $dS_{2\pm}$ & Saddle \\
 $S_{+}$ & Repellor \\
 $S_{-}$ & Attractor \\
 \hline
 \end{tabular}
 \caption{\label{tab:Linear_Stability} The linear stability character for the fixed points of the full system given in Table \ref{tab:Fixed_Points}.}
\end{table}

\begin{figure}
\begin{center}
\includegraphics[width=90mm]{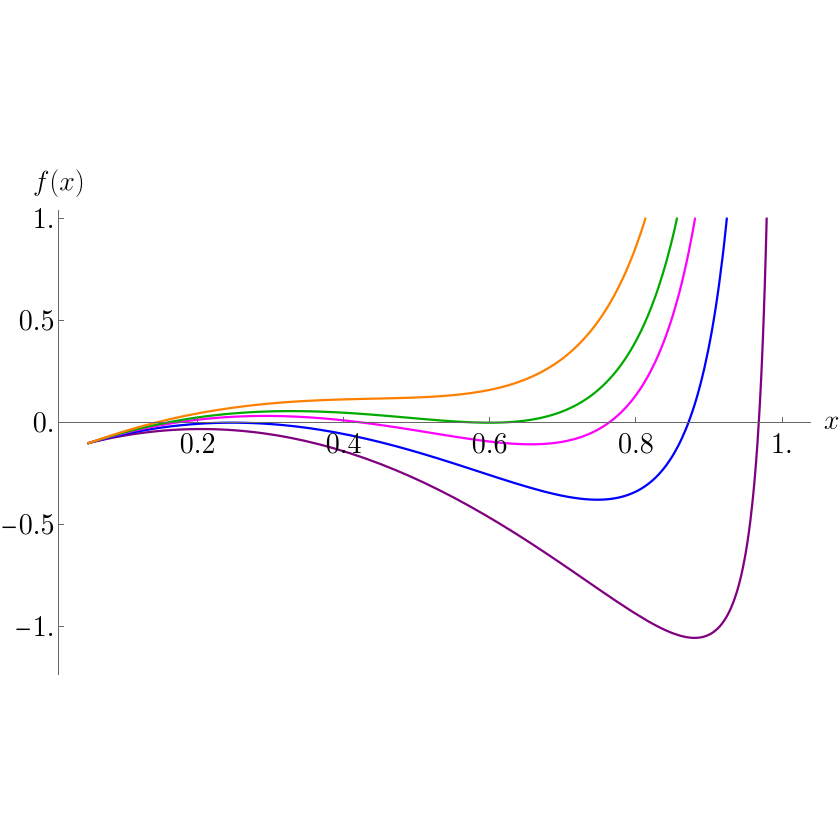}
\end{center}
\caption{A plot of $f(x)$ \eqref{EinsteinPoints} with varying values of $c_z$ to show the possible number of Einstein points admitted by the system. Starting from the top we have $c_z > 0.38$ (orange), $c_z \simeq 0.38$ (green), $0.20 < c_z < 0.38$ (magenta), $c_z \simeq 0.20$ (blue) and $c_z < 0.20$ (purple). The system always admits at least one Einstein point, and a maximum of three can exist.}
\label{fig:Wplot}
\end{figure}

At an Einstein point, the remaining expression left in the Hubble expansion variable \eqref{CompactHubbleRateDE}, which we call $f(x)$, is

\begin{equation}
    f(x) = z(x) + 2r(x) + x(1 + 3\scriptR) - 3\scriptR - 3x^2\,,
    \label{EinsteinPoints}
\end{equation}

\noindent which we need to solve numerically. An Einstein point exists whenever this expression is equal to zero. Taking the limit of $f(x)$ when $x \rightarrow \scriptR$ we find $f(x) \rightarrow -2\scriptR$, and taking the limit of $f(x)$ when $x \rightarrow 1$ we find $f(x) \rightarrow \infty$. Therefore, between $x = \scriptR$ and $x = 1$, $f(x)$ is always equal to zero at least once, and so at least one Einstein point always exists. The possible number of Einstein points admitted by the system can be seen in Fig. \ref{fig:Wplot}. For each curve in the plot, which have a specific value of $c_z$, an Einstein point occurs when the curve meets the $x$-axis at $f(x) = 0$. Between the smallest and largest values of $c_z$, we find the maximum number of Einstein points the system may admit is three.

The cases where two Einstein points are admitted are limiting cases, where two roots coincide, and where the maximum or minimum of the $f(x)$ curve touches but does not cross $f(x) = 0$. To find the values of $c_z$ in these limiting cases, we first need to find the derivative of $f(x)$ \eqref{EinsteinPoints} with respect to $x$, which is

\begin{multline}
    \frac{df(x)}{dx} = \frac{1}{3(1-x)(x-\scriptR)} [18x^3 - 3x^2(7+9\scriptR) \\
    + 3x(1 + 10\scriptR + 3\scriptR^2) - 3\scriptR(1+3\scriptR) + 3z + 8r]\,.
\end{multline}

\noindent Setting this equal to zero and rearranging for $c_z$, we find

\begin{multline}
    c_z = \frac{1}{3}\left(\frac{1-x}{x-\scriptR}\right)^\frac{1}{1-\scriptR}[-18x^3 +3x^2(7 + 9\scriptR) \\ 
    - 3x(1 + 10\scriptR + 3\scriptR^2) +3\scriptR(1 + 3\scriptR) - 8r]\,,
    \label{eqn:cz}
\end{multline}

\noindent which we substitute into $f(x) = 0$ \eqref{EinsteinPoints}. Solving this equation for $x$, we find $x \simeq 0.25$ and $x \simeq 0.60$, which from \eqref{eqn:cz} correspond to $c_z \simeq 0.20$ and $c_z \simeq 0.38$. These values are shown in Fig. \ref{fig:Wplot} by the blue and green curves, respectively.

In the general case where there are three Einstein points, there is a limiting case. In general, there are two separate CFS curves when three Einstein points are present; one through the lower and one through the upper Einstein point. In the limiting case, the two CFS curves coincide and the maxima of the potential are equal. To find where this occurs, we choose an initial value for $c_z$, then equate the potential U \eqref{eqn:U} at the upper and lower Einstein points. We iterate until we converge on a value of $c_z$, which occurs at $c_z \simeq 0.32$.

Therefore, we find seven different dynamical cases for this system. The open and flat models in each case all evolve in the same way. Expanding (contracting) models evolve from (to) an initial (a future) singularity to (from) a flat de-Sitter spacetime. Closed models within the FFS but outside the CFS also evolve in this way in each case, as the curvature is never large enough for a bounce or turn-around to occur. The behaviour of models within the CFS changes depending on the value of $c_z$ as this affects the number and character of the Einstein points present. We present these sub-cases in the following subsections, starting with the smallest range of $c_z$ and increasing the value with each case.

\subsection{1 Einstein Point}

\begin{figure}
\begin{center}
\includegraphics[width=80mm]{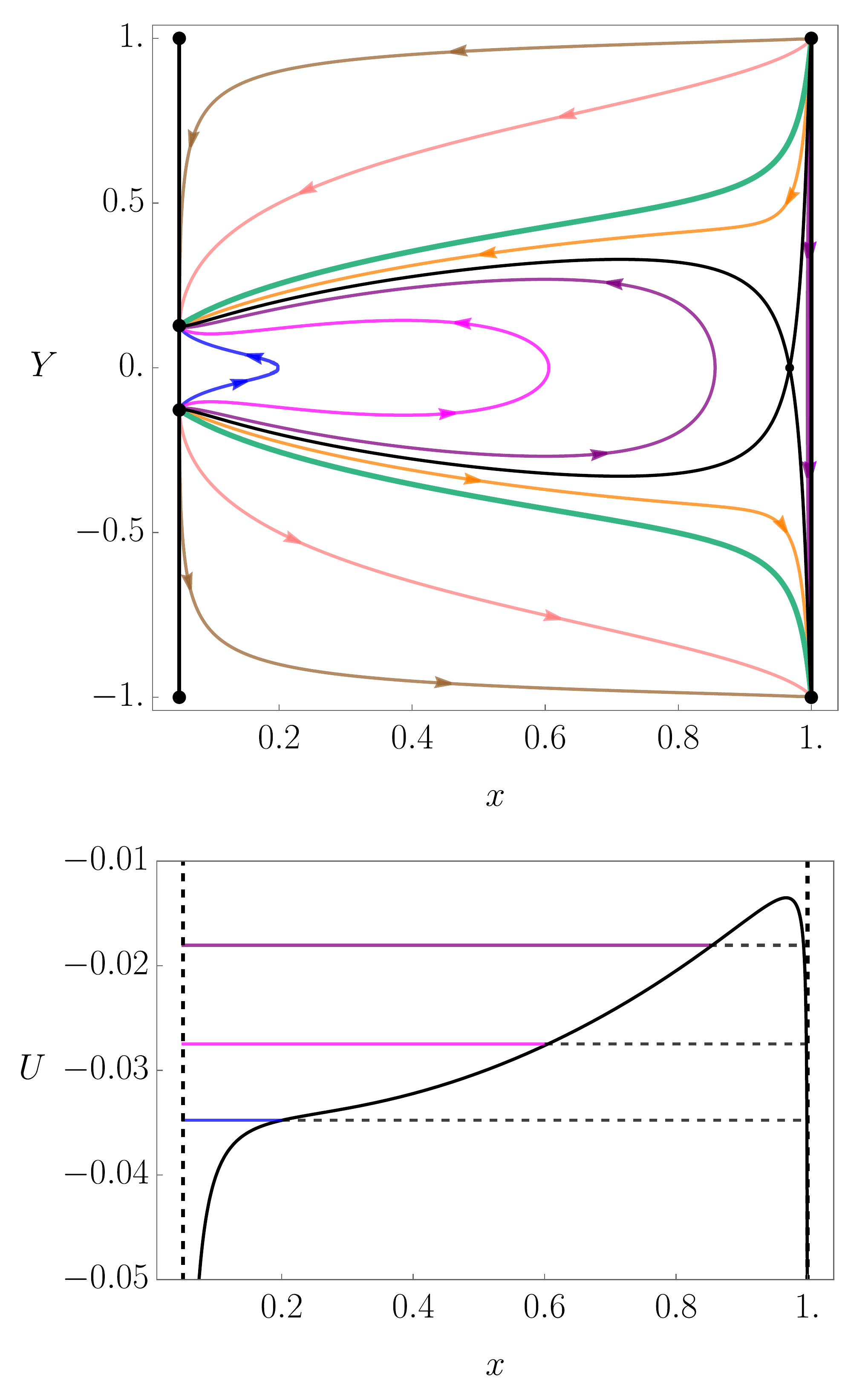}
\end{center}
\caption{$c_z < 0.20$ case, top panel: the projection of the full 4-D dynamics on the 2-D $x$-$Y$ plane. Bottom panel: the corresponding potential as in Eq. \eqref{eqn:U}, where $a$ is given by Eq. \eqref{eqn:a(x)}, $z$ by Eq. \eqref{eqn:z(x)} and $r$ by Eq. \eqref{eqn:r(x)}, and trajectories of the same colour in the two panels correspond to each other. One Einstein fixed point $x_E \simeq 0.97$ (a saddle) exists, which corresponds to the maximum of the potential, therefore we obtain bouncing trajectories for $x < 0.97$ within the CFS, however they are always accelerating.}
\label{fig:xYc1}
\end{figure}

The phase space for the range $c_z < 0.20$ is given in Fig \ref{fig:xYc1}. One Einstein point exists at $x_E \simeq 0.97$ which is a saddle point, and corresponds to the maximum of the potential. Most of the phase space within the CFS to the left of the Einstein point consists of bouncing models driven by the dark energy, while matter and radiation are subdominant, which evolve between two de Sitter fixed points. For the models which are dominated by matter and radiation, to the right of the Einstein point, the evolution is between two singularities with a turn-around. For all trajectories, the expansion (contraction) is always accelerating ($a'' > 0$) to the left of the Einstein point. Therefore, for this case, the bouncing models are always accelerating, and the turn-around models are never accelerating.

\subsection{2 Einstein Points}

\begin{figure}
\begin{center}
\includegraphics[width=80mm]{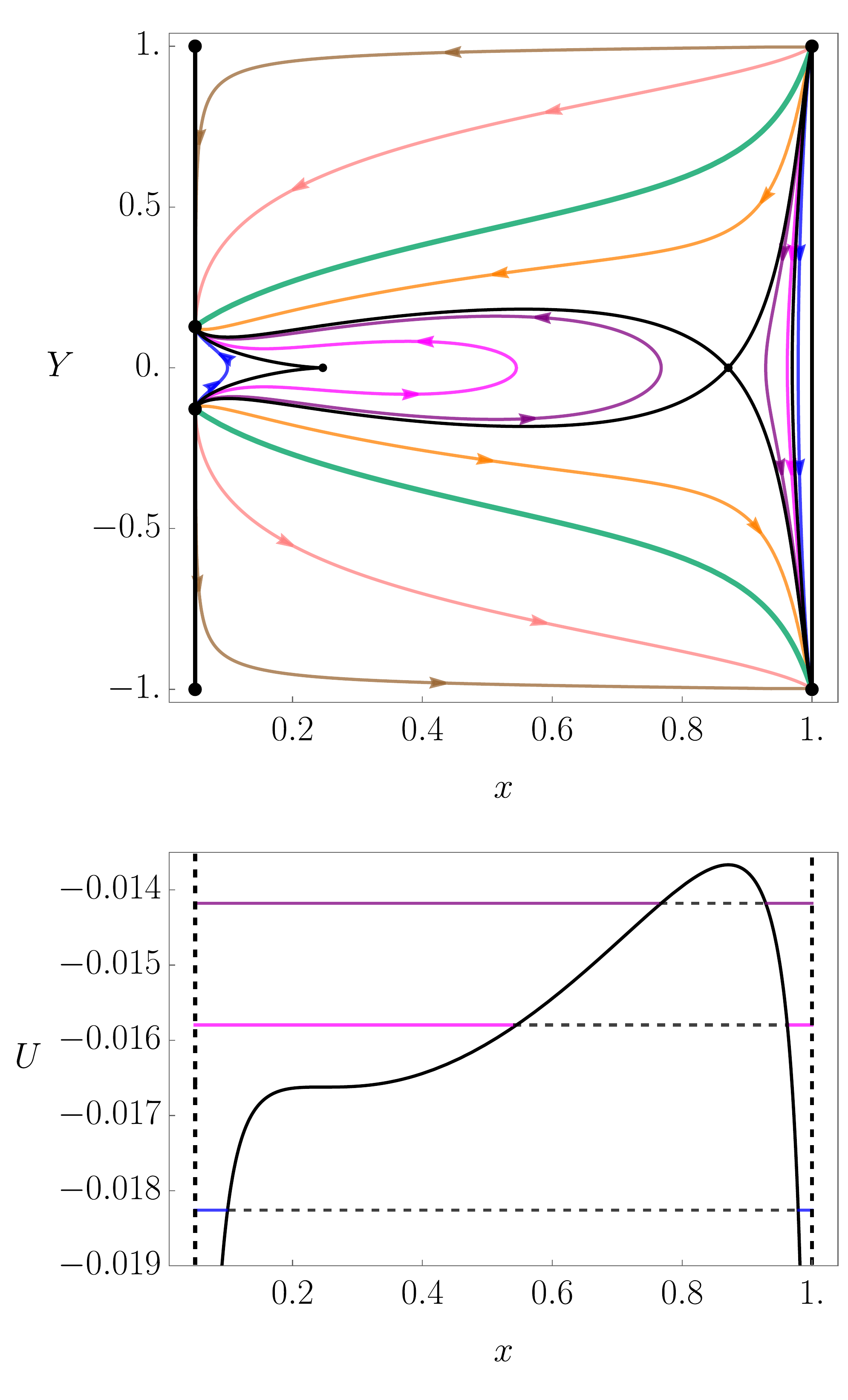}
\end{center}
\caption{The first limiting case where $c_z \simeq 0.20$, top panel: the projection of the full 4-D dynamics on the 2-D $x$-$Y$ plane. Bottom panel: the corresponding potential as in Eq. \eqref{eqn:U}, where $a$ is given by Eq. \eqref{eqn:a(x)}, $z$ by Eq. \eqref{eqn:z(x)} and $r$ by Eq. \eqref{eqn:r(x)}, and trajectories of the same colour in the two panels correspond to each other. Two Einstein fixed points occur. The fixed point $x_E \simeq 0.25$ (a cusp) corresponds to the horizontal point of inflection in the potential, and the fixed point at $x_E \simeq 0.87$ (a saddle) corresponds to the maximum of the potential. We obtain bouncing trajectories for $x < 0.87$ within the CFS, however these are always accelerating.}
\label{fig:xYc2}
\end{figure}

The phase space for the first limiting case with two Einstein points where $ c_z \simeq 0.20 $ is shown in Fig. \ref{fig:xYc2}. The Einstein point at $x_E \simeq 0.87$ is a saddle point, which corresponds to the maximum of the potential and is part of the outer CFS. Within this CFS, dark matter and radiation dominate for $x > 0.87$ where turn-around models evolve between two singularities. For $x < 0.87$, dark energy is dominant and we obtain bouncing models which evolve between two de Sitter points. The Einstein point at $x_E \simeq 0.25$ is a cusp (corresponding to the two de Sitter points coinciding), which corresponds to the horizontal point of inflection in the potential, and is part of the innermost CFS. Within this CFS, the bouncing models also evolve between the two de Sitter points. To the left of the Einstein point at $x_E \simeq 0.87$, trajectories are always accelerating, therefore the bouncing models are always accelerating, and the re-collapsing models are always decelerating.

\subsection{3 Einstein Points}

\subsubsection{Extra bouncing region}

\begin{figure}
\begin{center}
\includegraphics[width=80mm]{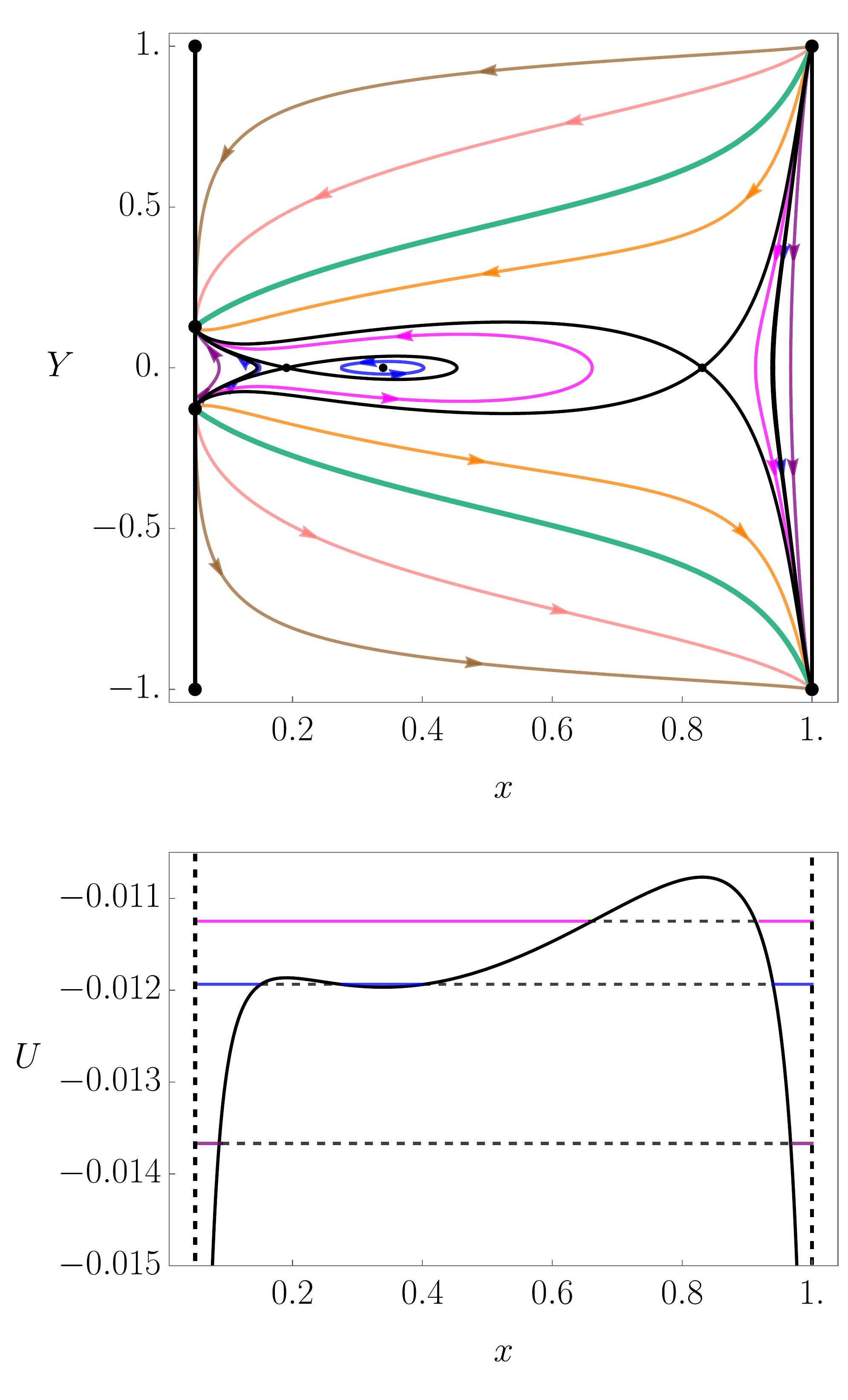}
\end{center}
\caption{$0.20 < c_z < 0.32$ case, top panel: the projection of the full 4-D dynamics on the 2-D $x$-$Y$ plane. Bottom panel: the corresponding potential as in Eq. \eqref{eqn:U}, where $a$ is given by Eq. \eqref{eqn:a(x)}, $z$ by Eq. \eqref{eqn:z(x)} and $r$ by Eq. \eqref{eqn:r(x)}, and trajectories of the same colour in the two panels correspond to each other. The Einstein fixed points $x_E \simeq 0.19$ and $x_E \simeq 0.83$ (saddle) correspond to the maxima of the potential, and $x_E \simeq 0.34$ (centre) to a local minimum. Bouncing behaviour is obtained for $x < 0.83$ within the outer CFS, and cyclic behaviour is obtained for $x > 0.19$ within the inner CFS. The cyclic models evolve with early acceleration and late-time deceleration, and the bouncing models evolve with early- and late-time acceleration connected by a decelerating period.}
\label{fig:xYc3}
\end{figure}

The first case where three Einstein points exist is shown in Fig. \ref{fig:xYc3}. The Einstein point at $x_E \simeq 0.19$ is a saddle point, which corresponds to a local maximum of the potential, and is part of the innermost CFS. Within this CFS, bouncing models evolve between two de Sitter points. Cyclic models are present around the Einstein fixed point at $x_E \simeq 0.34$, which is a centre, corresponding to a local minimum of the potential. These models contract until they bounce, and expand until they reach a turn-around. The Einstein point at $x_E \simeq 0.83$ is another saddle point, corresponding to the maximum of the potential, and is part of the outermost CFS. Within this CFS, turn-around models which evolve between two singularities are present for $x > 0.83$, where dark matter and radiation are dominant over the dark energy. For $x < 0.83$, dark energy is the dominant component, and bouncing models are present between the two CFS curves which evolve between two de Sitter points. These bounces evolve with an early- and late-time acceleration, connected by a period of deceleration. These are the physically important models, which we discuss in section \ref{sec:accn_regions}.

\subsubsection{Limiting case}

\begin{figure}
\begin{center}
\includegraphics[width=80mm]{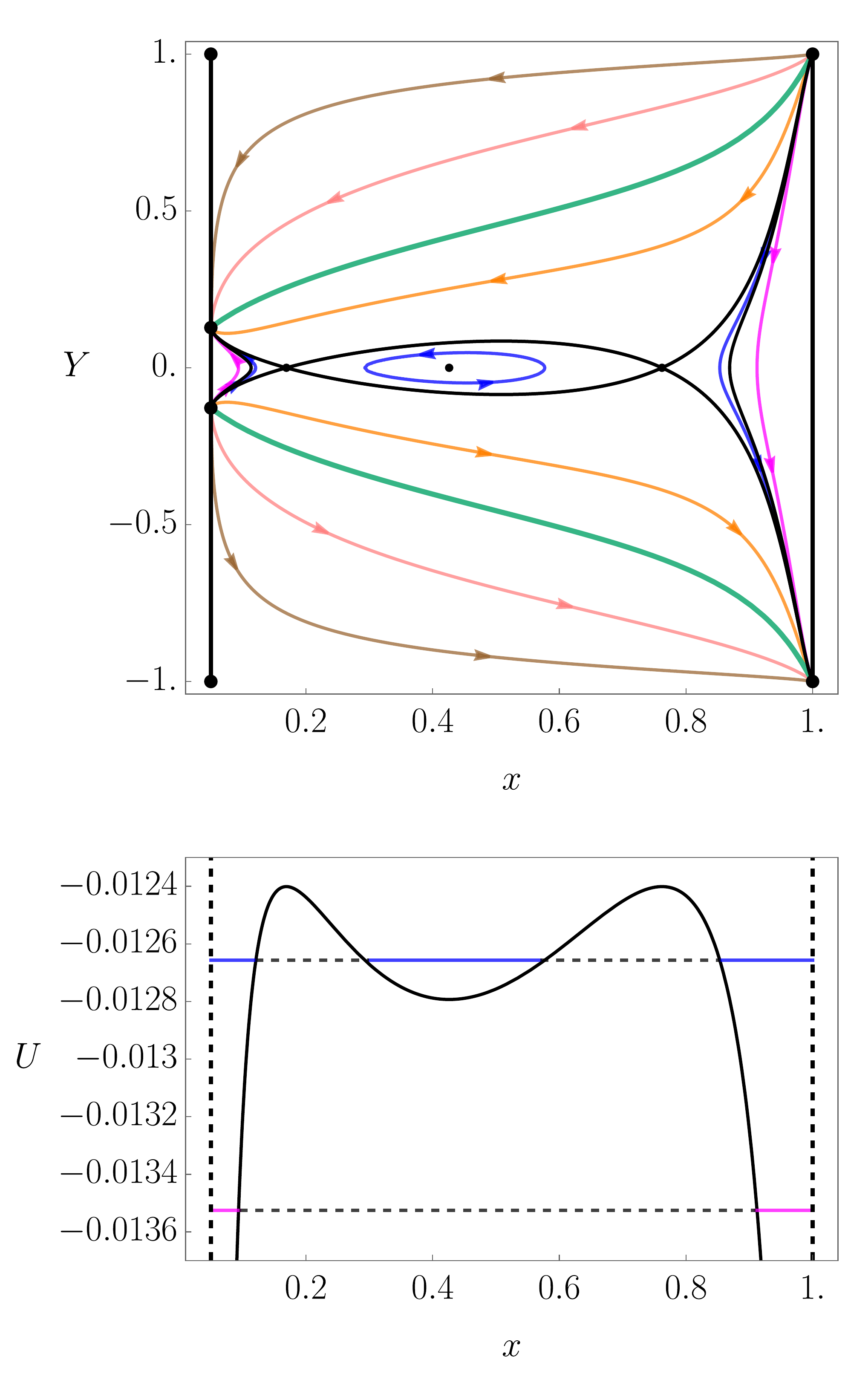}
\end{center}
\caption{The limiting case $c_z \simeq 0.32$, where the two CFS curves coincide. Top panel: the projection of the full 4-D dynamics on the 2-D $x$-$Y$ plane. Bottom panel: the corresponding potential as in Eq. \eqref{eqn:U}, where $a$ is given by Eq. \eqref{eqn:a(x)}, $z$ by Eq. \eqref{eqn:z(x)} and $r$ by Eq. \eqref{eqn:r(x)}, and trajectories of the same colour in the two panels correspond to each other. The fixed points $x_E \simeq 0.17$ and $x_E \simeq 0.76$ (saddle) correspond to the maxima of the potential which have the same value of the potential, and $x_E \simeq 0.43$ (centre) to a local minimum. Cyclic behaviour is obtained for $0.17 < x < 0.76$ and bouncing behaviour for $x < 0.17$, both within the CFS.}
\label{fig:xYc4}
\end{figure}

A limiting case exists for the general case with three Einstein points, when the two maxima of the potential have the same value. This case is shown in Fig. \ref{fig:xYc4}. Here, the two saddle Einstein fixed points at $x_E \simeq 0.17$ and $x_E \simeq 0.76$ correspond to the two maxima of the potential, and the two CFS curves corresponding to each of the fixed points merge and coincide. Within the CFS, turn-around models which evolve between two singularities are present in the $x > 0.76$ region of the phase space, where dark matter and radiation are dominant. Bouncing models which evolve between two de Sitter points are present when $x < 0.17$. Cyclic models are present around the Einstein point at $x_E \simeq 0.43$ (a centre) which corresponds to the local minimum value of the potential. These models contract until they bounce, and expand until they reach a turn-around. In this case, the cyclic models evolve with an early acceleration and late-time deceleration, the bouncing models always accelerate, and the re-collapsing models never accelerate.

\subsubsection{Extra turn-around region}

\begin{figure}
\begin{center}
\includegraphics[width=80mm]{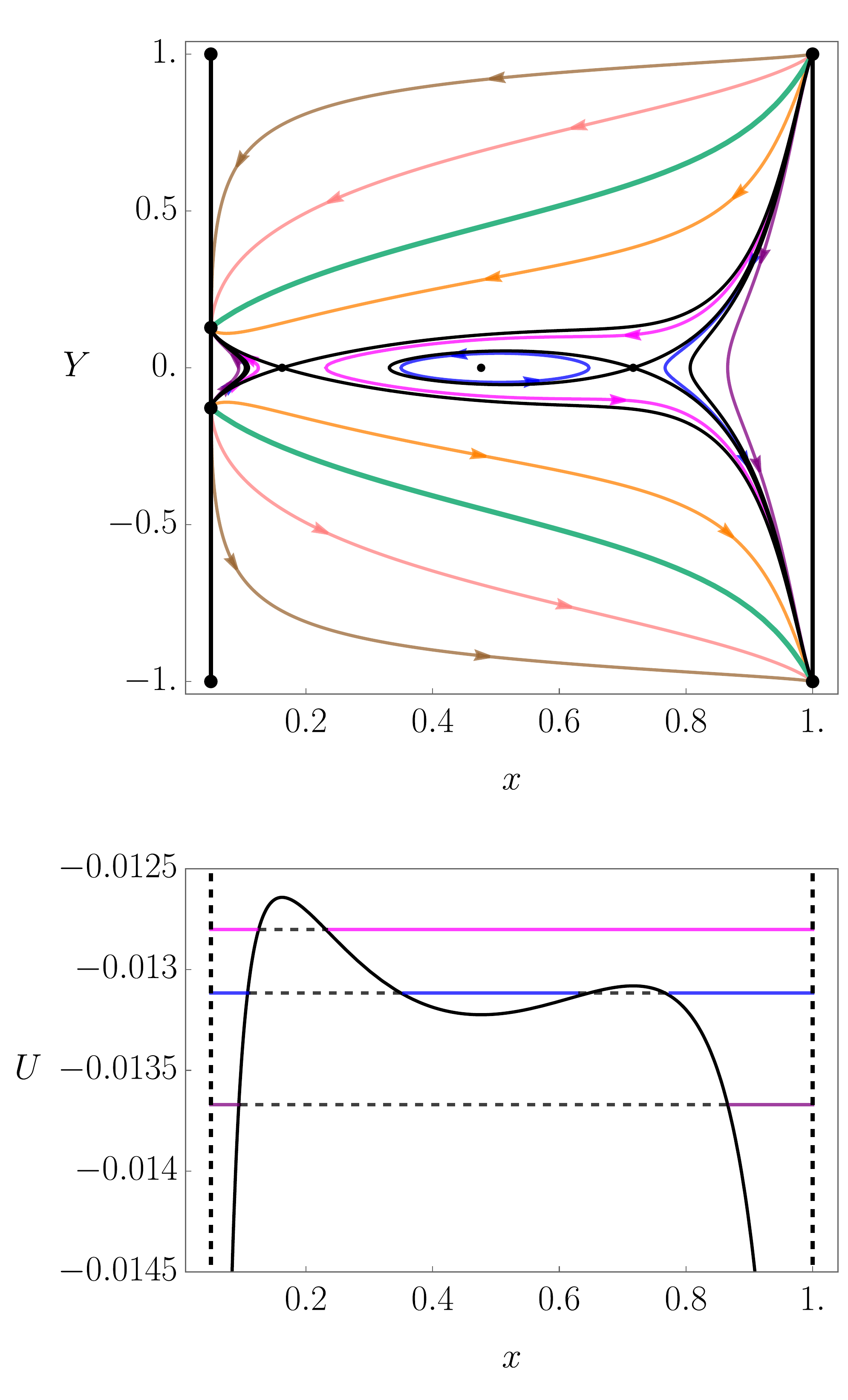}
\end{center}
\caption{$0.32 < c_z < 0.38$ case, top panel: the projection of the full 4-D dynamics on the 2-D $x$-$Y$ plane. Bottom panel: the corresponding potential as in Eq. \eqref{eqn:U}, where $a$ is given by Eq. \eqref{eqn:a(x)}, $z$ by Eq. \eqref{eqn:z(x)} and $r$ by Eq. \eqref{eqn:r(x)}, and trajectories of the same colour in the two panels correspond to each other. The fixed points $x_E \simeq 0.16$ and $x_E \simeq 0.72$ (saddle) correspond to the maxima of the potential, and $x_E \simeq 0.48$ (centre) to a local minimum. Cyclic behaviour is obtained for $x < 0.72$ within the inner CFS and bouncing behaviour for $x < 0.16$ within the outer CFS. Cyclic models evolve with early acceleration and late-time deceleration, and the bouncing models always accelerate.}
\label{fig:xYc5}
\end{figure}

The final case with three Einstein points is given in Fig. \ref{fig:xYc5}. The Einstein point at $x_E \simeq 0.72$ is a saddle point, which corresponds to the local maximum value of the potential, and is part of the innermost CFS. Within this CFS, turn-around models which evolve between two singularities are present for the $x > 0.72$ region of the phase space. These models never accelerate, and dark matter and radiation are dominant. For $x < 0.72$, dark energy is the dominant component. Cyclic models are present around the centre Einstein fixed point at $x_E \simeq 0.48$, which has a local minimum value of the potential. These models contract until they bounce, then expand until they turn-around and evolve with early acceleration and late-time deceleration. The Einstein fixed point at $x_E \simeq 0.16$ is another saddle point which is part of the outermost CFS, and occurs at the maximum of the potential. Bouncing models evolve between two de Sitter points for $x < 0.16$, which always accelerate. For the $x > 0.16$ region of the phase space between the two CFS curves, turn-around models are present which evolve between two singularities. In these models, radiation and dark matter are initially dominant, but decrease more rapidly than the dark energy, which becomes dominant as the trajectory passes the fixed point at $x_E \simeq 0.72$. After the turn-around, the radiation and matter increase more rapidly than the dark energy, and become dominant again when $x > 0.72$, where the models approach a singularity. These re-collapsing models evolve with an early and late-time deceleration connected by an accelerating period.

\subsection{2 Einstein Points}

\begin{figure}
\begin{center}
\includegraphics[width=80mm]{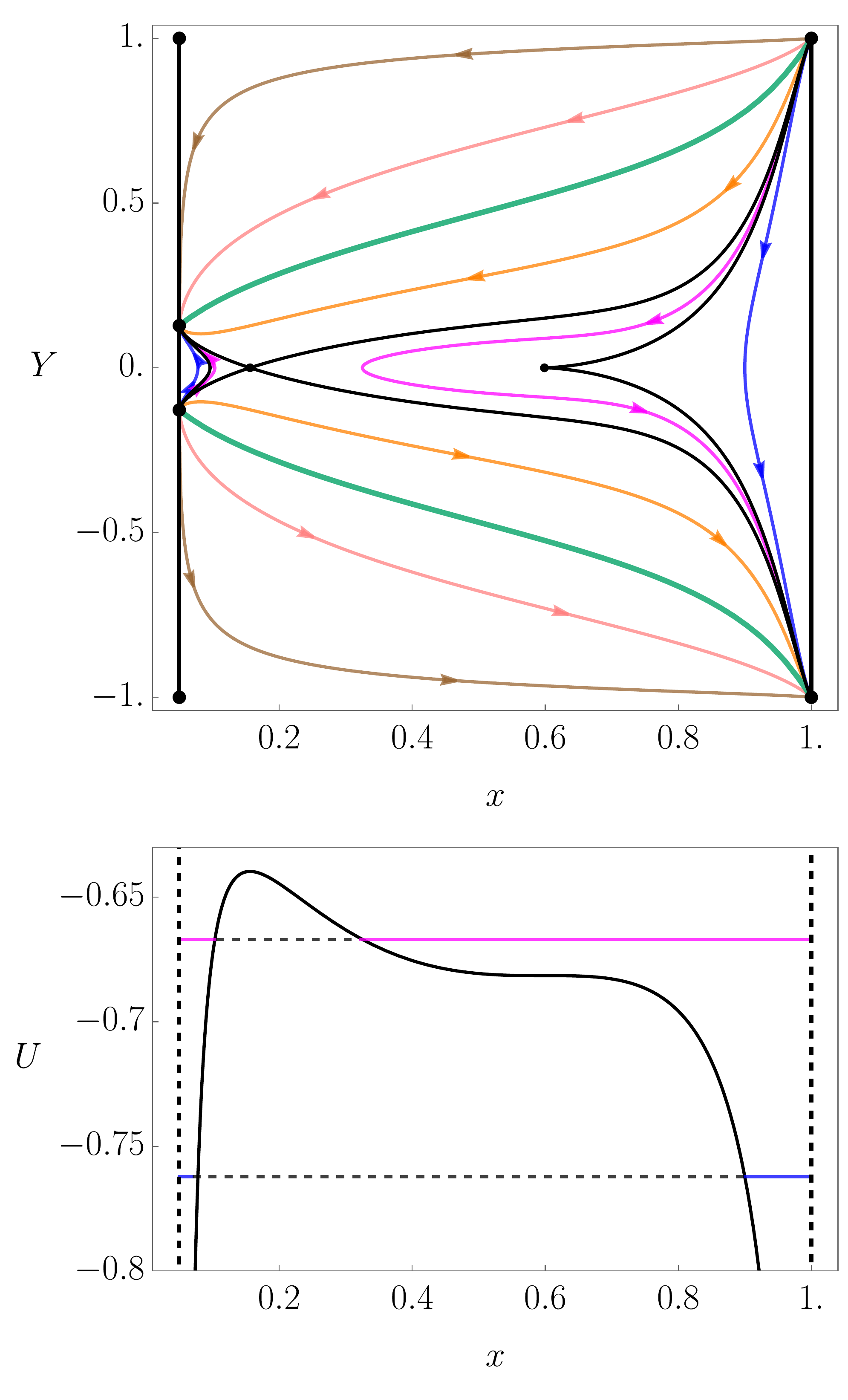}
\end{center}
\caption{$c_z \simeq 0.38$ case, top panel: the projection of the full 4-D dynamics on the 2-D $x$-$Y$ plane. Bottom panel: the corresponding potential as in Eq. \eqref{eqn:U}, where $a$ is given by Eq. \eqref{eqn:a(x)}, $z$ by Eq. \eqref{eqn:z(x)} and $r$ by Eq. \eqref{eqn:r(x)}, and trajectories of the same colour in the two panels correspond to each other. The fixed point $x_E \simeq 0.16$ (saddle) corresponds to the maximum of the potential, and $x_E \simeq 0.60$ (cusp) to the horizontal point of inflection in the potential. Bouncing behaviour is obtained within the CFS for $x < 0.16$, however these models always accelerate.}
\label{fig:xYc6}
\end{figure}

The phase space for the system for $c_z \simeq 0.38$ is shown in Fig. \ref{fig:xYc6}, which is the second limiting case with two Einstein points. The Einstein point at $x_E \simeq 0.60$ is a cusp (corresponding to the two de Sitter points coinciding), which corresponds to the horizontal point of inflection in the potential, and is part of the innermost CFS. Within this CFS, turn-around models evolve between two singularities, where the dark matter and radiation are the dominant components, and so never accelerate. The Einstein point at $x_E \simeq 0.16$ is a saddle which is part of the outermost CFS, and corresponds to the maximum value of the potential. For the $x < 0.16$ region of the phase space, bouncing models evolve between two de Sitter points. In these models, the dark energy always dominates, and they always accelerate. Between the two CFS curves, turn-around models are present which evolve between two singularities, and never accelerate. Dark matter and radiation are dominant for $x > 0.60$, and dark energy is dominant for the $x < 0.60$ region of the phase space. After the turn-around, the radiation and dark matter components increase more quickly than the dark energy throughout the collapse, and become dominant on approach to the singularity.

\subsection{1 Einstein Point} \label{subsection:1_E_Pt}

\begin{figure}
\begin{center}
\includegraphics[width=80mm]{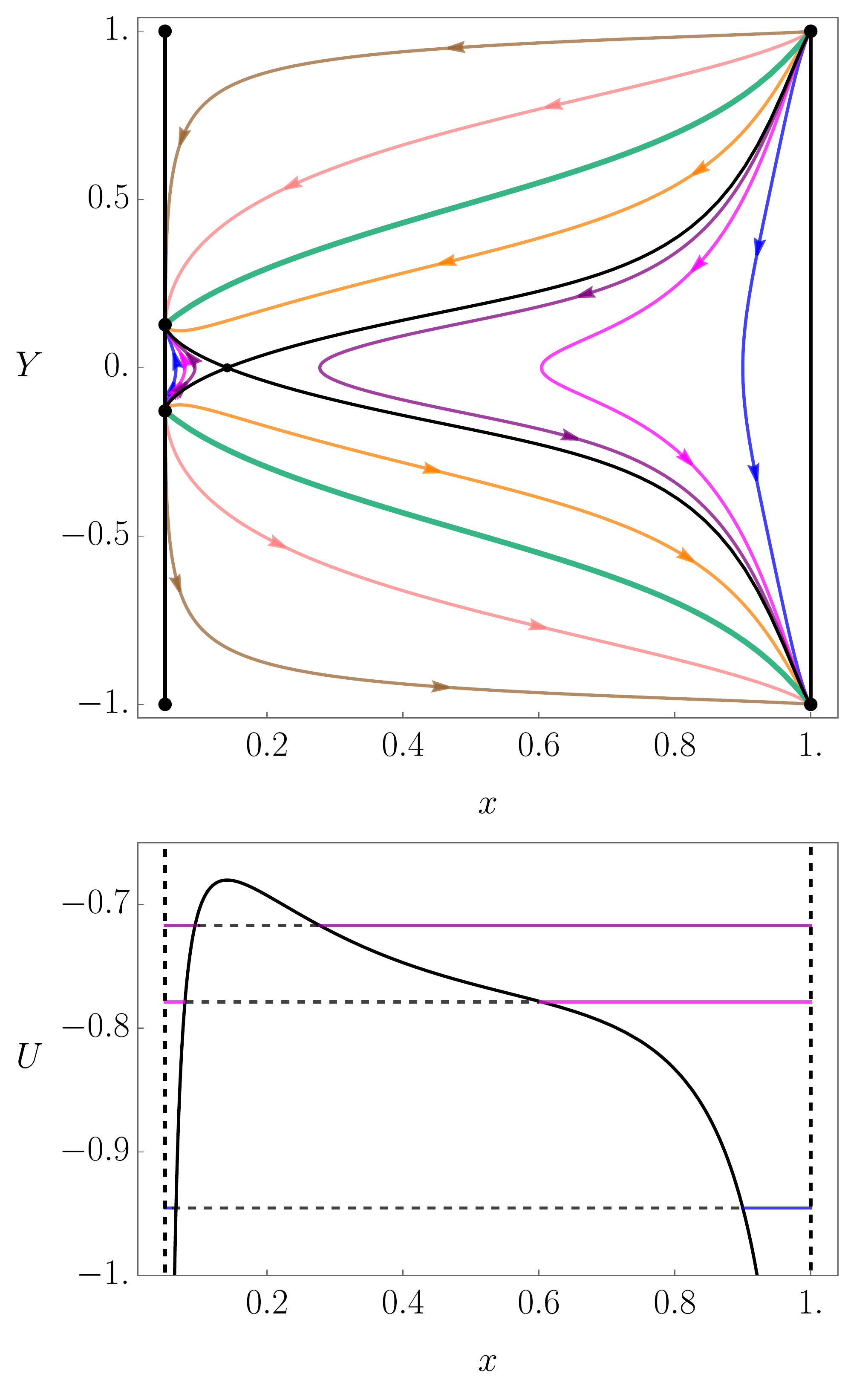}
\end{center}
\caption{$c_z > 0.38$ case, top panel: the projection of the full 4-D dynamics on the 2-D $x$-$Y$ plane. Bottom panel: the corresponding potential as in Eq. \eqref{eqn:U}, where $a$ is given by Eq. \eqref{eqn:a(x)}, $z$ by Eq. \eqref{eqn:z(x)} and $r$ by Eq. \eqref{eqn:r(x)}, and trajectories of the same colour in the two panels correspond to each other. The fixed point $x_E \simeq 0.14$ (saddle) corresponds to the maximum of the potential. Bouncing models are obtained within the CFS for $x < 0.14$, which always accelerate.}
\label{fig:xYc7}
\end{figure}

The final case for this system is given in Fig. \ref{fig:xYc7}. Qualitatively, the behaviour here is the same as in Fig. \ref{fig:xYc1}, except that the Einstein point in this case is much closer to $x = \scriptR$ at $x_E \simeq 0.14$. This is a saddle point that corresponds to the maximum value of the potential for the system, and is part of the CFS. Within the CFS, bouncing models are present for the $x < 0.14$ region of the phase space, in which dark energy is always dominant. For the $x > 0.14$ region, turn-around models which evolve between two singularities exist. These models are initially dominated by matter and radiation, and then by dark energy. After the turn-around, the dark matter and radiation increase more quickly than the dark energy, and become dominant again as the singularity is approached. For all trajectories, the expansion (contraction) is always accelerating to the left of the Einstein point, therefore the bouncing models always accelerate, and the re-collapse models always decelerate.

\subsection{Acceleration regions} \label{sec:accn_regions}

In order to determine which of the cases are viable, we can consider the acceleration regions in each phase space. For our models to be feasible, they must have an early and late time acceleration, connected by a decelerating phase where large-scale structure can form. The acceleration equation for this system is given by

\begin{equation}
    \frac{a''}{a} = z + 2r - 3\scriptR + (1 + 3\scriptR)x - 3x^2\,.
    \label{eqn:accn}
\end{equation}

\begin{figure}
\begin{center}
\includegraphics[width=80mm]{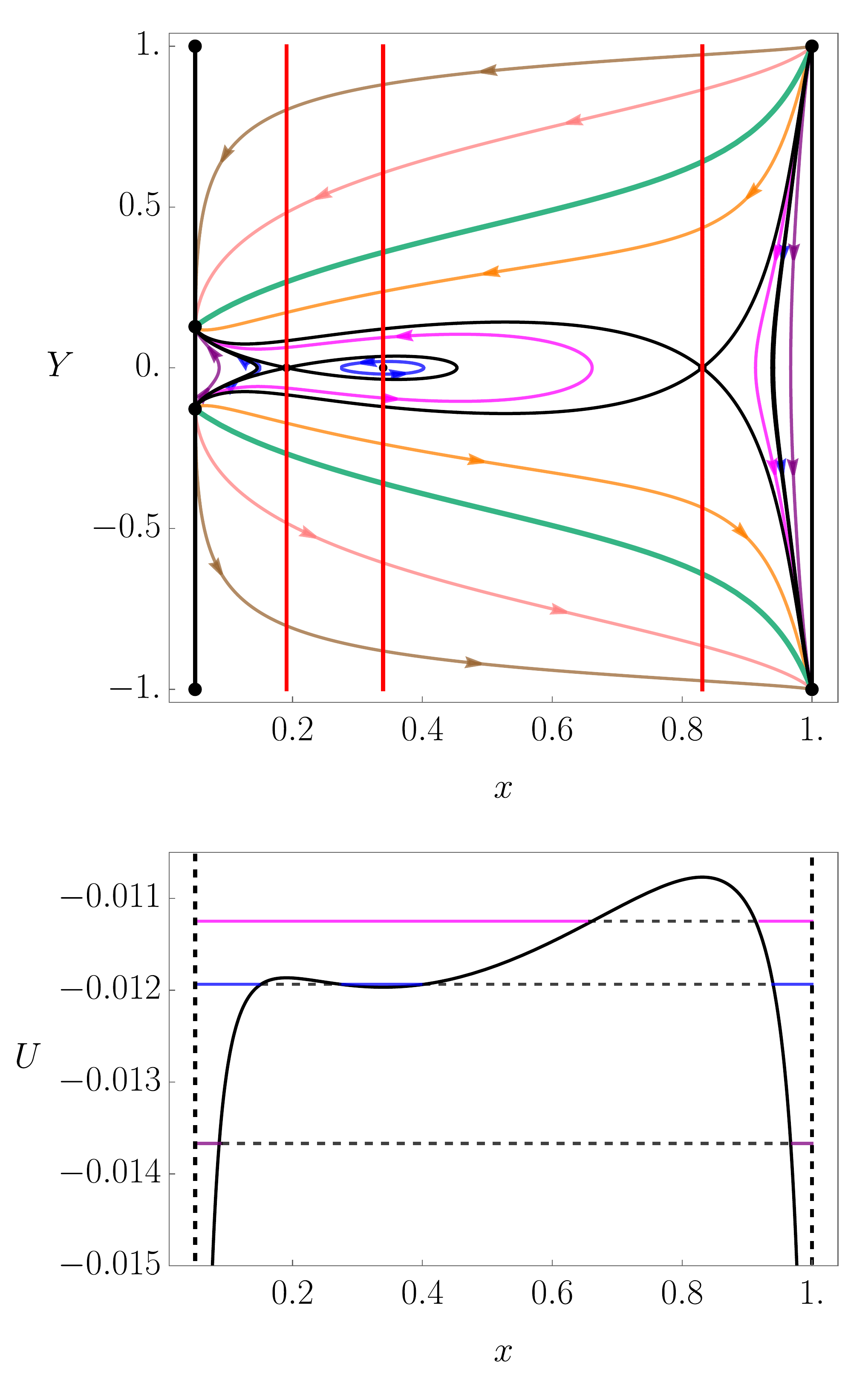}
\end{center}
\caption{The phase space (top panel) and corresponding potential (bottom panel) for $0.20 < c_z < 0.32$ as in Fig. \ref{fig:xYc3}, with the boundaries between accelerating and decelerating regions in red. The $0 < x < 0.19$ and $0.34 < x < 0.83$ regions are accelerating, and the $0.19 < x < 0.34$ and $0.83 < x < 1$ regions are decelerating.}
\label{fig:xYc3_Accn}
\end{figure}

\noindent We can find the boundaries between accelerating and decelerating regions by setting the acceleration equation \eqref{eqn:accn} to zero. We find that the boundaries where acceleration is zero go through each Einstein fixed point parallel the the $Y$-axis. Therefore, one case remains where a bouncing model exhibits early and late time acceleration, which is the $0.20 < c_z < 0.32$ case. This case is shown again in Fig. \ref{fig:xYc3_Accn}, where we have included the red $a'' = 0$ boundaries. The $0 < x < 0.19$ and $0.34 < x < 0.83$ regions are accelerating, and the $0.19 < x < 0.34$ and $0.83 < x < 1$ regions are decelerating. Therefore, the bouncing models present between the two CFS curves evolve with an early- and late-time accelerated phase, and are therefore the models of interest. 

However, we find that for all of the cases we have presented, the bouncing behaviour is spoiled when dark matter and radiation become dominant, and turn-around models occur for trajectories in the phase space near $x = 1$. Therefore, these bouncing models are only relevant when matter and radiation appear after the bounce, and would need a mechanism such as inflation with a subsequent period of reheating in order to be feasible. These models also only occur when they have sufficient curvature. Although we cannot make a quantitative statement from our qualitative analysis, in principle, these models would be in tension with observations \cite{Planck2018VI}.

\section{Upper bounds on $z$ and $r$} \label{sec:FullSystemUpperBound}

In light of the results thus far, we introduce an upper bound on the dark matter and radiation, such that their energy densities cannot reach infinity. The idea is that matter and radiation are not always present in standard cosmology, but that they are created in a period of reheating. It is beyond the scope of our paper to introduce a reheating phase, perhaps through an interaction between radiation, dark matter and the dark energy component, so we simply assume that matter and radiation have an upper bound. In the following, we focus on investigating whether bouncing models can be the general closed model when matter and radiation appear after the bounce has occurred, even for models that are close to being spatially flat. 

In order to implement an upper bound on matter and radiation, we assume that their EoS's are modified by a non-linear term at high energies in a similar way to the high energy bound for the dark energy component. Physically, these EoS's are not particularly meaningful, however this is a simple way of imposing an energy scale at which matter and radiation appear such that the bounce is not spoiled. As far as our qualitative analysis is concerned, the specific physical mechanism for having an upper bound for matter and radiation is not that relevant. We therefore re-write Eq. \eqref{darkmatterDE} and Eq. \eqref{radiationDE} as 

\begin{equation}
    z' = -\frac{3Yz(1 - \frac{z}{z_*})}{\sqrt{1-Y^2}}
    \label{eqn:z'_upper_bound}
\end{equation}

\begin{equation}
    r' = -\frac{4Yr(1 - \frac{r}{r_*})}{\sqrt{1-Y^2}}\,,
    \label{eqn:r'_upper_bound}
\end{equation}

\noindent where $z_*$ and $r_*$ are the characteristic energy scales of dark matter and radiation, respectively. Should we model a proper inflationary era, we could set $z_*$ and $r_*$ at the energy scale of reheating. Here, we will simply assume that the upper bounds imposed on matter and radiation are at the highest possible energy scale, coinciding with the high energy cosmological constant of the dark energy, somewhere between the Planck scale and that of inflation. This assumption does not qualitatively change the dynamics, and is the most robust way of investigating the survival of the bounce, therefore we set $z_* = r_* = 1$. 

The effective EoS's for dark matter and radiation can be found by equating $z'$ \eqref{eqn:z'_upper_bound} and $r'$ \eqref{eqn:r'_upper_bound} to the continuity equation. We find

\begin{equation}
    w_z = -z
    \label{eqn:wz_upper_bound}
\end{equation}

\begin{equation}
    w_r = \frac{1}{3} - \frac{4}{3}r\,.
    \label{eqn:wr_upper_bound}
\end{equation}

\noindent For $z > 1/3$ and $r > 1/2$, we find their respective effective EoS's are less than -1/3, and so the dark matter and radiation can contribute to acceleration in these regions of the phase space. We find that the Raychaudhuri equation for $Y$, the compact variable representing the Hubble expansion scalar, now becomes

\begin{multline}
    Y' = -Y^2(1-Y^2)^\frac{1}{2} - \frac{(1-Y^2)^\frac{3}{2}}{6}[z(1-3z) \\
    + 2r(1-2r) - 3\scriptR + x(1+3\scriptR) -3x^2]\,,
    \label{eqn:Y'_Upper_Limits}
\end{multline}
 
\noindent and $x'$ is still as in Eq. \eqref{CompactDarkEnergyDE}. As previously, we project the 4-D dynamics onto the $x$-$Y$ plane. Integrating $z$ \eqref{eqn:z'_upper_bound} with respect to $x$ \eqref{CompactDarkEnergyDE}, we find

\begin{equation}
    z(x) = \frac{c_z\left(\frac{x-\scriptR}{1-x}\right)^\frac{1}{1-\scriptR}}{1+c_z\left(\frac{x-\scriptR}{1-x}\right)^\frac{1}{1-\scriptR}}\,,
    \label{eqn:zofx_upper_bound}
\end{equation}

\noindent and integrating $r$ \eqref{eqn:r'_upper_bound} with respect to $x$ \eqref{CompactDarkEnergyDE}, we obtain

\begin{equation}
    r(x) = \frac{c_r\left(\frac{x-\scriptR}{1-x}\right)^\frac{4}{3(1-\scriptR)}}{1 + c_r\left(\frac{x-\scriptR}{1-x}\right)^\frac{4}{3(1-\scriptR)}}\,.
    \label{eqn:rofx_upper_bound}
\end{equation}

\noindent We can then calculate the first integral $c_r$ in same way as before. Using the Planck 2018 density parameters \cite{Planck2018VI}, along with our definition of $\scriptR$ \eqref{eqn:dimensionless_variables} and the low energy cosmological constant $\rho_\Lambda$ \eqref{eqn:rho_Lambda}, we find

\begin{equation}
    c_r = \frac{\scriptR \Omega_r}{\alpha \Omega_\Lambda - \scriptR \Omega_r}\left(\frac{\alpha - \scriptR}{\scriptR (1 - \alpha)}\right)^\frac{4}{3(1-\scriptR)}\simeq 0.00077
    \label{eqn:cr_upper_bound}
\end{equation}

\noindent for a value of $\scriptR = 0.05$.

The fixed points for this system are given in Table \ref{tab:Fixed_Points_UB}. In this system, there are no physical singularities, but the fixed points $dS_{2\pm}$ and $dS_{4 \pm}$ are coordinate singularities of the de Sitter spacetime when represented as an FLRW, see \cite{Hawking1973}. 

The eigenvalues of the fixed points for this system are shown in Table \ref{tab:Eigenvalues_UB}, and the linear stability classification for each point is given in Table \ref{tab:Linear_Stability_UB}. As previously, we cannot include the eigenvalues of the fixed points at $Y = \pm 1$, however we find from the Raychaudhuri equation  \eqref{eqn:Y'_Upper_Limits} that the de Sitter point at $Y = + 1$ is a repeller and the de Sitter point at $Y = -1$ is an attractor along $x = 1$. The de Sitter points along $x = \scriptR$ at $Y = \pm 1$ are saddle points, with trajectories moving away from the $Y = +1$ point and towards the $Y = -1$ point. Similarly to before, the stability character of the Einstein fixed points depends on the value of the first integral $c_z$ for a fixed $\scriptR$ and $c_r$. Similarly to before, the stability character of the Einstein fixed points depends on the value of the first integral $c_z$ for a fixed $\scriptR$ and $c_r$.


In this case, in order to find the Einstein points from the Raychaudhuri equation \eqref{eqn:Y'_Upper_Limits}, we need to find the zeros of the following function

\begin{multline}
    f(x) = z(x)[1-3z(x)] + 2r(x)[1-2r(x)] \\ - 3\scriptR + x(1+3\scriptR) - 3x^2 \,.
    \label{eqn:w_upper_bound}
\end{multline}

\onecolumngrid

\begin{table}
\centering
\setlength{\tabcolsep}{5pt}
 \begin{tabular}{||c | c c c c||} 
 \hline
 Name & $x$ & $Y$ & $z$ & $r$\\ [0.5ex] 
 \hline\hline
 $E$ & $x$ & $0$ & $z$ & $\frac{1}{4}\left(1 - \sqrt{1 + 4(x + z) + 12(x\scriptR - \scriptR - x^2 - z^2)}\right)$\\
 dS$_{1\pm}$ & $\scriptR$ & $\pm\sqrt{\frac{\scriptR}{3 + \scriptR}}$ & $0$ & $0$ \\
 dS$_{2\pm}$ & $\scriptR$ & $\pm 1$ & $0$ & $0$ \\
 dS$_{3\pm}$ & $1$ & $\pm \frac{1}{\sqrt{2}}$ & $1$ & $1$\\
 dS$_{4\pm}$ & $1$ & $\pm 1$ & $1$ & $1$ \\
 \hline
 \end{tabular}
 \caption{\label{tab:Fixed_Points_UB} The fixed points of the system when $z$ and $r$ have an upper bound}
\end{table}

\onecolumngrid
         
\begin{table}[h!]
\centering
 \begin{tabular}{||c | c c c c||} 
 \hline
  Name & $\lambda_1$ & $\lambda_2$ & $\lambda_3$ & $\lambda_4$ \\ 
 \hline\hline
 $E$ &  0 & 0 & \thead{$\frac{1}{\sqrt{6}} \{-1 + 18x^3 -z + 9z^2(2z-1) + 27\scriptR - 9x^2(1 + 3\scriptR) - x $ \\ $+ [1 + 2z(3z - 1) + 2x(3x - 1)]\sqrt{1 + 4z(1-3z) - 12\scriptR + 4x(1-3x + 3\scriptR)}\}^\frac{1}{2}$} & $-\lambda_3$ \\
 $dS_{1\pm}$ &  $\mp 4 \sqrt{\frac{\scriptR}{3}}$ & $\mp \sqrt{3\scriptR}$ & $\mp 2 \sqrt{\frac{\scriptR}{3}}$ & $\mp \sqrt{3\scriptR}(1 - \scriptR)$ \\
 $dS_{3\pm}$ & $\pm 4$ & $\pm 3$ & $\mp 2$ & $\pm 3 (1 - \scriptR)$ \\
 \hline
 \end{tabular}
 \caption{\label{tab:Eigenvalues_UB} The eigenvalues for the fixed points of the system when $z$ and $r$ have an upper bound}
\end{table}

\vspace{1cm}

\twocolumngrid

\begin{table}[h!]
\centering
 \begin{tabular}{||c | c ||} 
 \hline
  Name & Stability Character \\ 
 \hline\hline
 $E$ & Centre, Saddle or Cusp \\
 $dS_{1+}$ & Attractor \\
 $dS_{1-}$ & Repellor \\
 $dS_{2\pm}$ & Saddle \\
 $dS_{3\pm}$ & Saddle \\
 $dS_{4+}$ & Repellor \\
 $dS_{4-}$ & Attractor \\
 \hline
 \end{tabular}
 \caption{\label{tab:Linear_Stability_UB} The linear stability character for the fixed points of the system when $z$ and $r$ have an upper bound}
\end{table}

\begin{figure}
\begin{center}
\includegraphics[width=90mm]{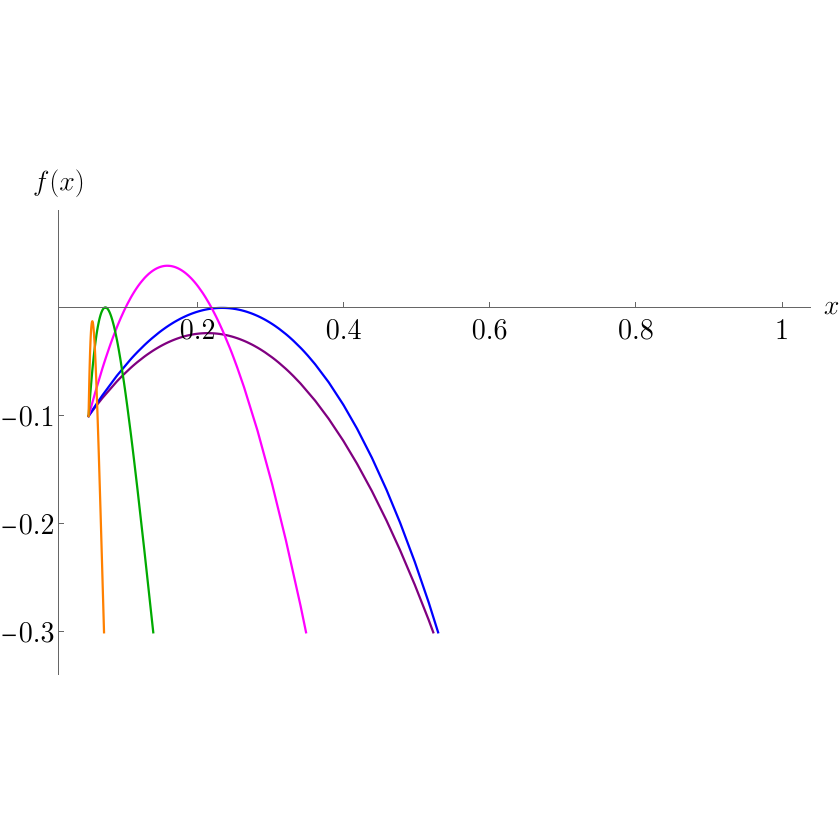}
\end{center}
\caption{$f(x)$ with varying values of $c_z$ to show the different number of Einstein points which can be admitted by the system. Starting from the left we have $c_z > 11.0$ (orange), $c_z \simeq 11.0$ (green), $0.25 < c_z < 11.0$ (magenta), $c_z \simeq 0.25$ (blue) and $c_z < 0.25$ (purple). When $c_z < 0.25$ or $c_z > 11.0$ no Einstein points are admitted. One Einstein point is admitted in the limiting cases when $c_z \simeq 0.25$ and $c_z \simeq 11.0$, and two Einstein points are admitted when $0.25 < c_z < 11.0$.}
\label{fig:Wplot_Upper_Bound}
\end{figure}

\noindent We need to solve numerically for the values $x_E$, such that $f(x_E) = 0$. Taking the limit of $f(x)$ when $x \rightarrow \scriptR$, we find $f(x) \rightarrow -2\scriptR$, and when $x \rightarrow 1$, we find $f(x) \rightarrow -6$. Therefore, this system does not necessarily admit any Einstein points between $x = \scriptR$ and $x = 1$ as both limits are negative. Fig. \ref{fig:Wplot_Upper_Bound} shows the range of Einstein points which can be admitted by the system for different values of $c_z$.

In total, we find three different cases for the dynamics, except here more than one range of $c_z$ can give the same qualitative dynamics. Increasing $c_z$ increases the contribution of the $z(1-3z)$ term to $f(x)$ \eqref{eqn:w_upper_bound} until $z = 1/3$. Once this point is reached, increasing $c_z$ then increases a negative contribution from the $z(1 - 3z)$ term. There are two limiting cases where one Einstein point is admitted which have the same qualitative dynamics that are shown by the green and blue curves in Fig. \ref{fig:Wplot_Upper_Bound}. As before, to find the values of $c_z$ in the limiting cases, we take the first derivative of $f(x)$ and solve for $c_z$ when $df(x)/dx = f(x) = 0$, and find $c_z \simeq 0.25$ and $c_z \simeq 11.0$. Both $c_z < 0.25$ and $c_z > 11.0$ admit the same qualitative dynamics, where no Einstein points are admitted. In the range $0.25 < c_z < 11.0$ two Einstein points are admitted.

The open and flat models evolve in the same way for each case. Expanding (contracting) open models evolve between two de Sitter fixed points, from (to) an open geometry to (from) a flat geometry. Expanding (contracting) flat models evolve between two de Sitter fixed points along the FFS. The dynamical behaviour of the closed models changes depending on the value of $c_z$. In each case, we find that the closed models avoid a singularity as $U \rightarrow 0$ as $x \rightarrow 1$, forming a potential barrier. Therefore, there are no turn-around models that evolve between singularities, and all closed models admit a bounce. We present these cases in the following subsections, and note that qualitatively the dynamics is the same as the sub-manifolds in Section \ref{sec:Submanifolds} when only dark energy is present.

\subsection{No Einstein points}

\begin{figure}
\begin{center}
\includegraphics[width=80mm]{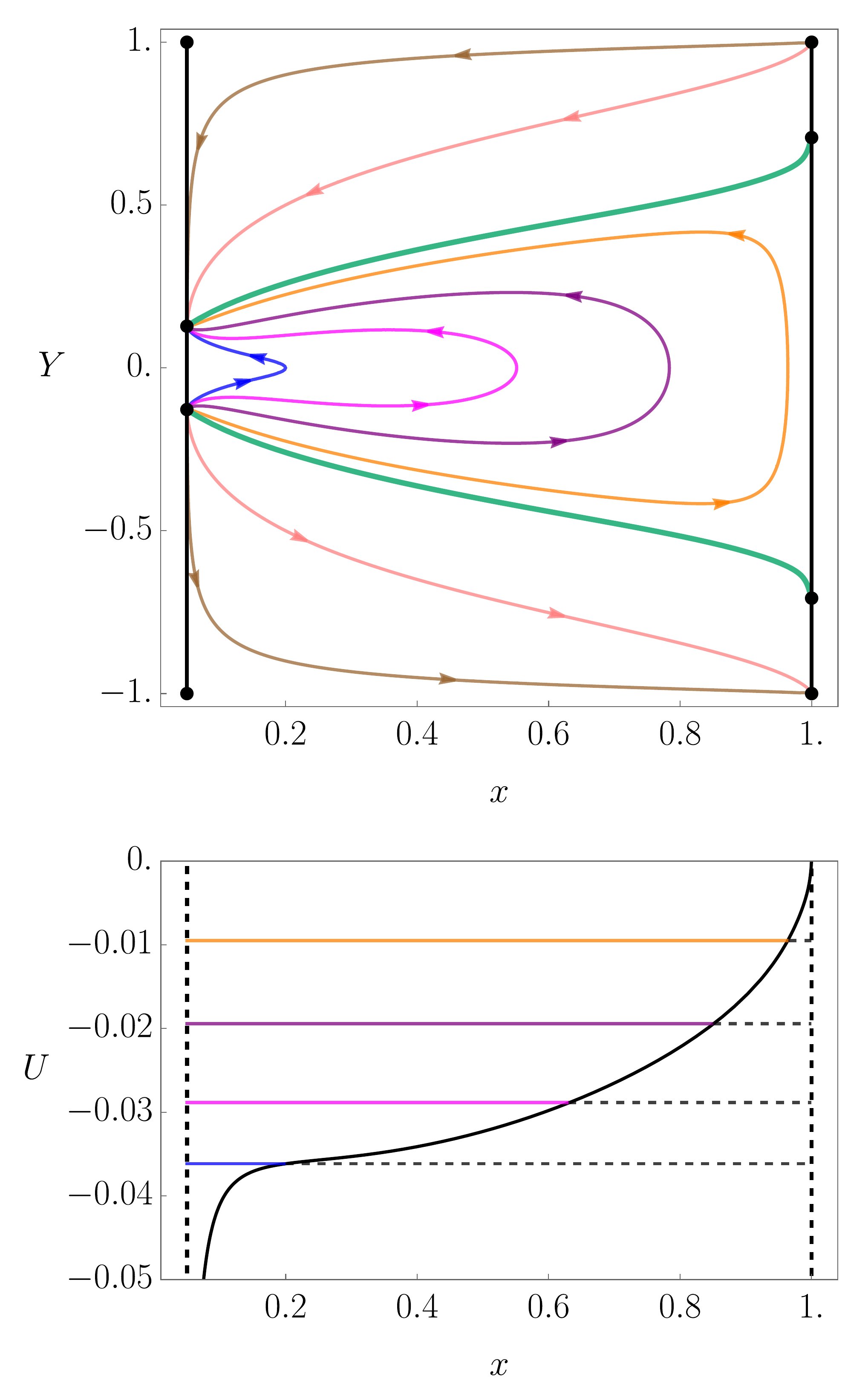}
\end{center}
\caption{$c_z < 0.25$ case, which is qualitatively the same as the $c_z > 11.0$ case. Top panel: the projection of the full 4-D dynamics on the 2-D $x$-$Y$ plane. Bottom panel: the corresponding potential as in Eq. \eqref{eqn:U}, where $a$ is given by Eq. \eqref{eqn:a(x)}, $z$ by Eq. \eqref{eqn:zofx_upper_bound} and $r$ by Eq. \eqref{eqn:rofx_upper_bound}. Trajectories of the same colour in the two panels correspond to each other. No Einstein points are admitted, and all closed models bounce, which always accelerate.}
\label{fig:xYc1_Upper_Bound}
\end{figure}

The phase space for the system when $c_z < 0.25$ is shown in Fig. \ref{fig:xYc1_Upper_Bound}, which is qualitatively the same as when $c_z > 11.0$. As before, the green outermost thick curve is the flat Friedmann Separatrix (FFS), which here has the general form

\begin{equation}
    \frac{Y^2}{1-Y^2} = \frac{x}{3} + \frac{z}{3} + \frac{r}{3} \,.
\end{equation}

\noindent The inner most curve is the closed Friedmann Separatrix (CFS), which has the general form

\begin{equation}
    \frac{Y^2}{1-Y^2} = \frac{x}{3} + \frac{z}{3} + \frac{r}{3} - \frac{1}{a^2}\left( \frac{x_E}{3} + \frac{z_E}{3} + \frac{r_E}{3} \right) \,.
\end{equation}

\noindent $z$ and $r$ are given by Eq. \eqref{eqn:zofx_upper_bound} and Eq. \eqref{eqn:rofx_upper_bound}, respectively, and $a$ is given by Eq. \eqref{eqn:a(x)}. As previously, the horizontal lines along $x = \scriptR$ and $x = 1$ are the de Sitter lines where $x' = z' = r' = 0$.

In this case, no Einstein points are admitted, and all closed models bounce, evolving between two de Sitter points. Therefore, when matter and radiation have an upper bound, the bounce is no longer spoiled and a singularity is avoided. When $c_z < 0.25$, dark energy is the dominant component for most of the phase space, until $x = 1$ is approached where dark matter and radiation become comparable to the dark energy. When $c_z > 11.0$, dark matter is the dominant component for most of the phase space, except sufficiently close to $x = 1$ where all three components are comparable, and close to $x = \scriptR$ where dark energy becomes dominant. All trajectories are always accelerating in this case.

\begin{figure}[h]
\begin{center}
\includegraphics[width=80mm]{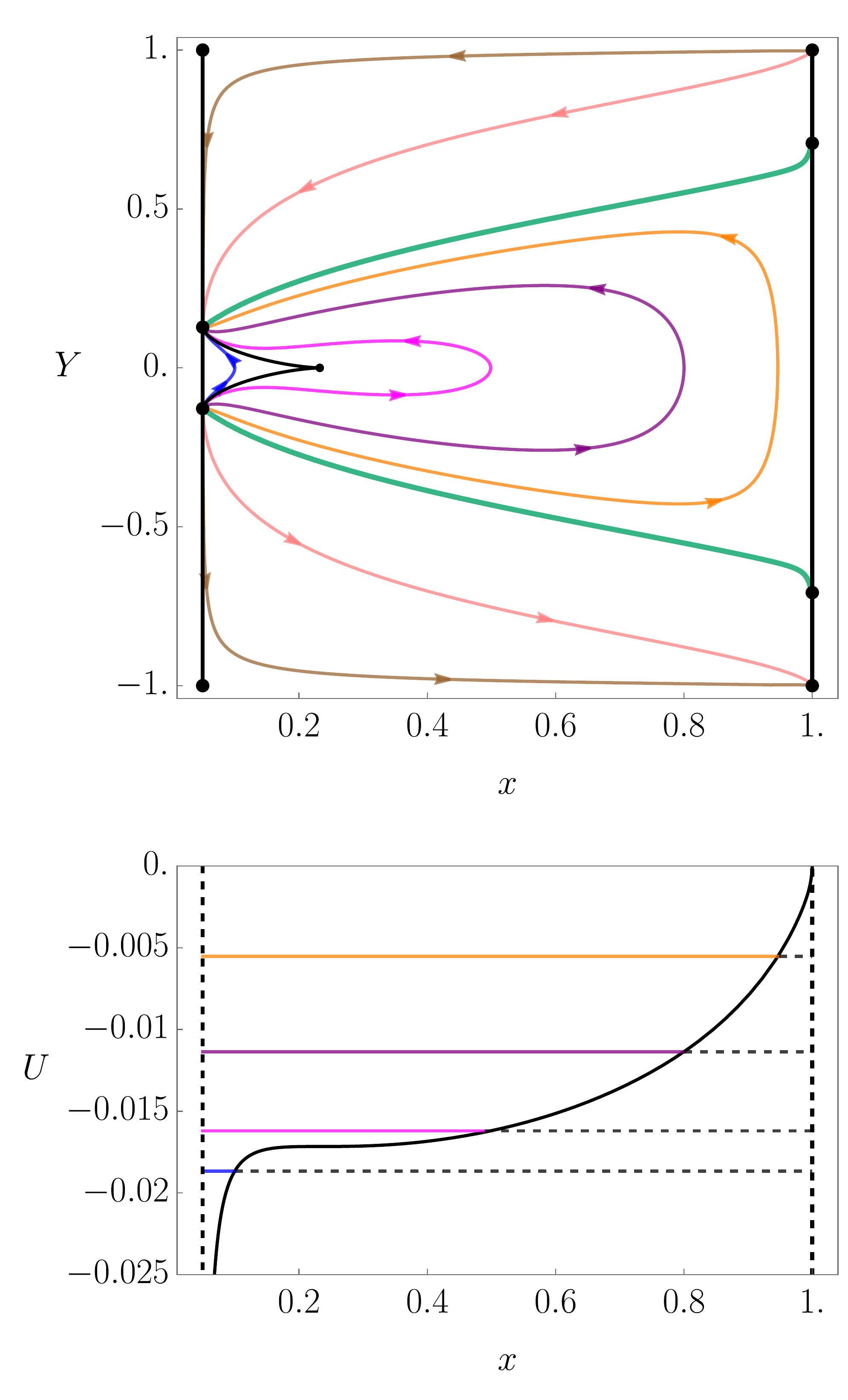}
\end{center}
\caption{$c_z \simeq 0.25$ case, which is qualitatively the same as the $c_z \simeq 11.0$ case. Top panel: the projection of the full 4-D dynamics on the 2-D $x$-$Y$ plane. Bottom panel: the corresponding potential as in Eq. \eqref{eqn:U}, where $a$ is given by Eq. \eqref{eqn:a(x)}, $z$ by Eq. \eqref{eqn:zofx_upper_bound} and $r$ by Eq. \eqref{eqn:rofx_upper_bound}. Trajectories of the same colour in the two panels correspond to each other. One Einstein point exists at $x_E \simeq 0.23$ (cusp) corresponding to the horizontal point of inflection in the potential. All closed models bounce, and these are always accelerating.}
\label{fig:xYc2_Upper_Bound}
\end{figure}

\subsection{One Einstein point}

The limiting case where $c_z \simeq 0.25$ and one Einstein point is admitted is shown in Fig. \ref{fig:xYc2_Upper_Bound}. The qualitative dynamics in this case is the same when $c_z \simeq 11.0$. One Einstein point is admitted at $x_E \simeq 0.23$ (cusp) which is part of the CFS, and corresponds to the horizontal point of inflection in the potential. All closed models bounce, evolving between two de Sitter fixed points. When $c_z \simeq 0.25$, dark energy is dominant for most of the phase space, and when $c_z \simeq 11$, dark matter is mostly dominant, except close to $x = \scriptR$ where dark energy is dominant. All three components become comparable near $x = 1$. Here, all trajectories are always accelerating.

\begin{figure}
\begin{center}
\includegraphics[width=80mm]{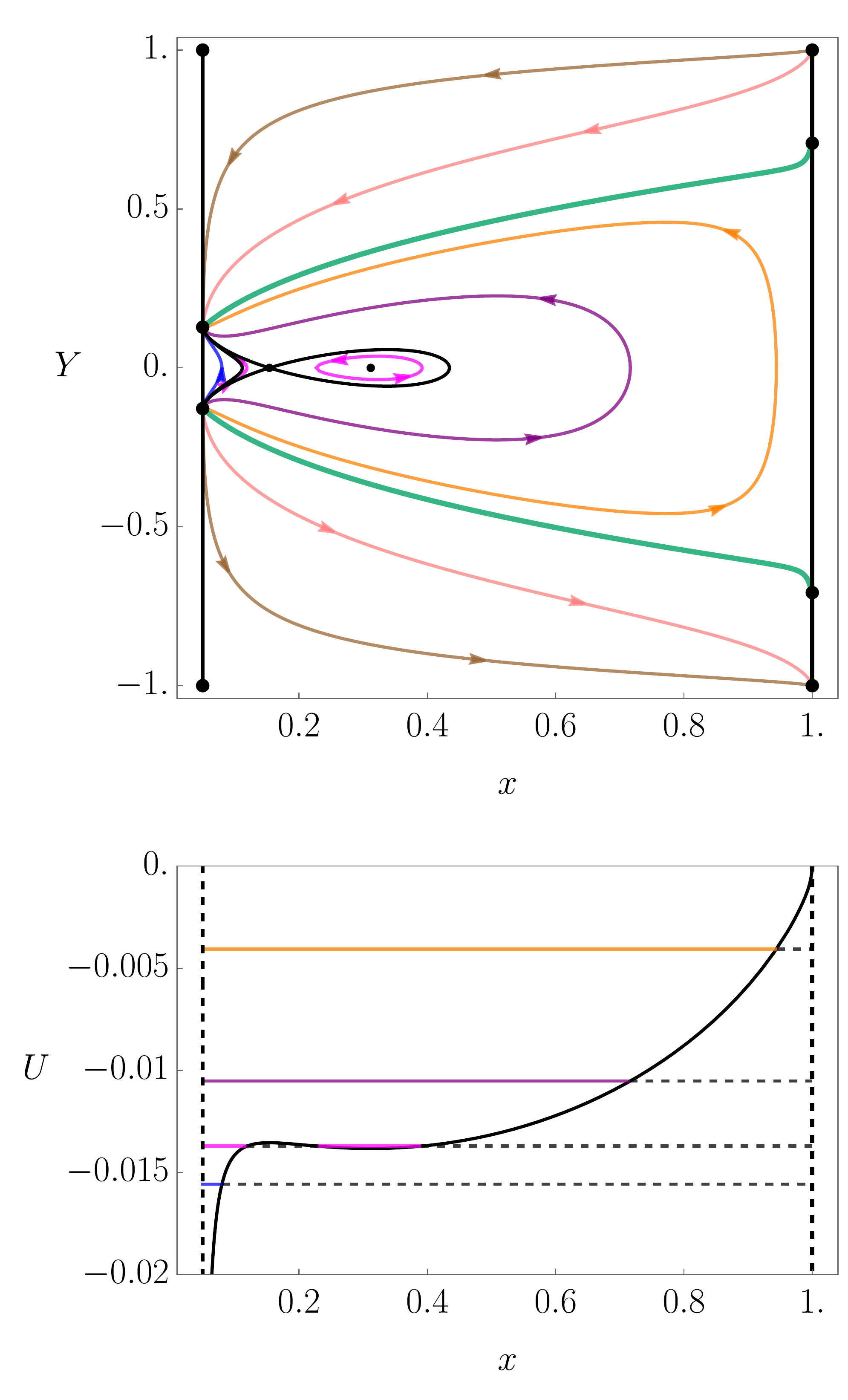}
\end{center}
\caption{$0.25 < c_z < 11.0$ case, top panel: the projection of the full 4-D dynamics on the 2-D $x$-$Y$ plane. Bottom panel: the corresponding potential as in Eq. \eqref{eqn:U}, where $a$ is given by Eq. \eqref{eqn:a(x)}, $z$ by Eq. \eqref{eqn:zofx_upper_bound} and $r$ by Eq. \eqref{eqn:rofx_upper_bound}. Trajectories of the same colour in the two panels correspond to each other. Two Einstein points exist at $x_E \simeq 0.15$ (saddle), which corresponds to a local maximum of the potential, and at $x_E \simeq 0.31$ (centre) corresponding to a local minimum of the potential. Trajectories around the $x_E \simeq 0.31$ fixed point within the CFS are cyclic, and all other closed models bounce. Bouncing models outside the CFS evolve with an early- and late-time acceleration, connected by a period of deceleration.}
\label{fig:xYc3_Upper_Bound}
\end{figure}

\subsection{Two Einstein points}

The case where two Einstein points are admitted is shown in Fig. \ref{fig:xYc3_Upper_Bound}. The Einstein point at $x_E \simeq 0.15$ (saddle) corresponds to a local maximum of the potential, and is part of the CFS. There are bouncing models which evolve between two de Sitter points within the CFS, and cyclic models around the Einstein point at $x_E \simeq 0.31$ (centre), which corresponds to a local minimum of the potential. These bouncing models always accelerate, and the cyclic models evolve with an early-time acceleration and a late-time deceleration. Outside of the CFS, all closed models bounce, evolving between two de Sitter points. These models evolve with an early- and late-time accelerated expansion, connected by a decelerating period. If the value of $c_z$ is closer to 0.25, most of the phase space will be dark energy dominated, however if its value is closer to 11.0, then dark matter will be mostly dominant.

\begin{figure}
\begin{center}
\includegraphics[width=80mm]{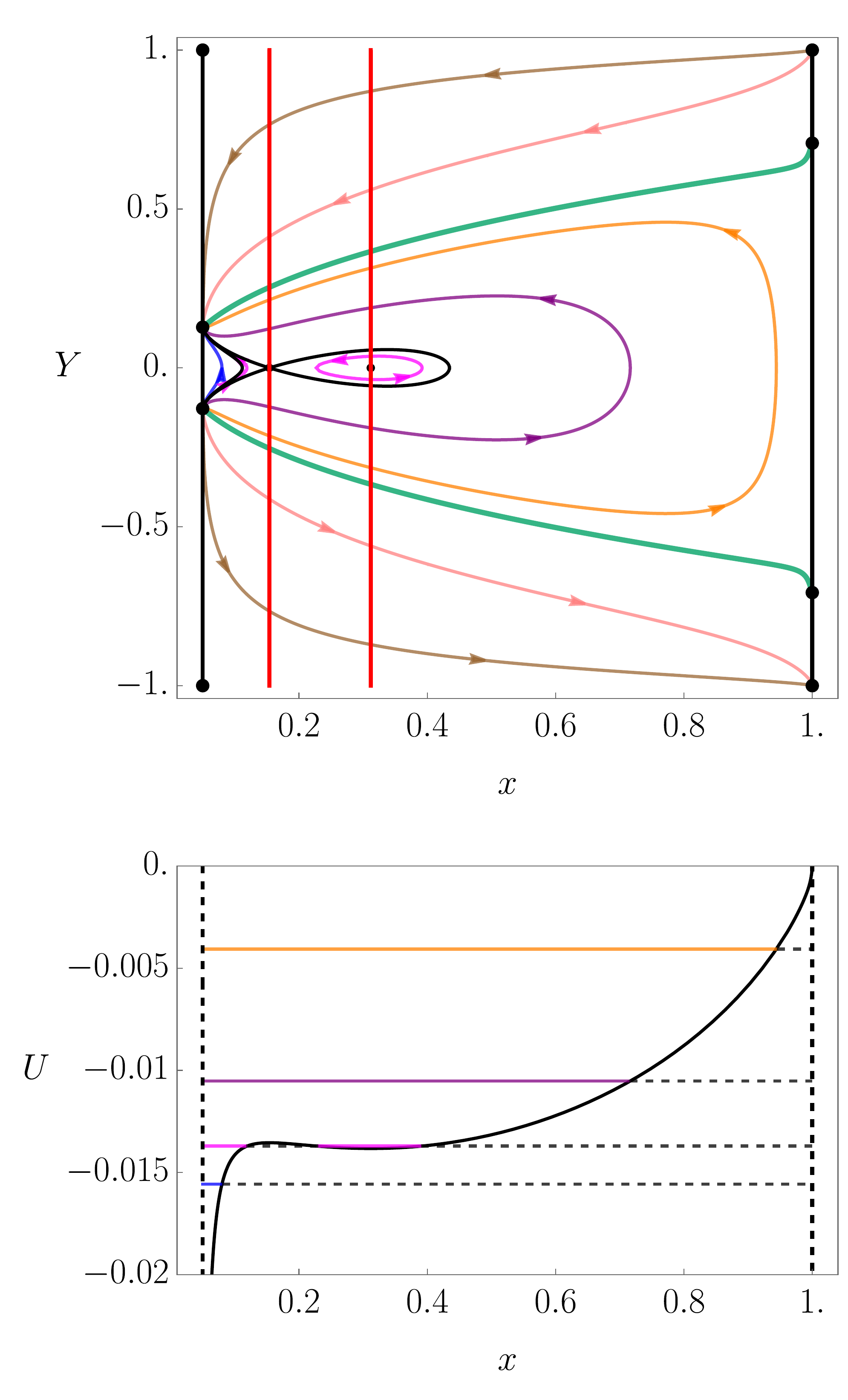}
\end{center}
\caption{The phase space (top panel) and corresponding potential (bottom panel) for $0.25 < c_z < 11.0$ as in Fig. \ref{fig:xYc3_Upper_Bound}, with the boundaries between accelerating and decelerating regions in red. The $0.15 < x < 0.31$ region of the phase space is decelerating, and the $x < 0.15$ and $x > 0.31$ regions are accelerating.}
\label{fig:xYc3_Upper_Bound_Accn}
\end{figure}

\subsection{Acceleration}

As before, we calculate where the acceleration is zero to find the boundaries between accelerating and decelerating regions of the phase space. In this system, the acceleration equation is given by
\begin{equation}
    \frac{a''}{a} = z(1-3z) + 2r(1-2r) - 3\scriptR + (1 + 3\scriptR)x - 3x^2\,.
    \label{eqn:accnupperbound}
\end{equation}

\noindent The case of interest in this new system is $0.25 < c_z < 11.0$, in which two Einstein points are admitted. This case with the boundaries between accelerating and decelerating regions of the phase space can be seen in Fig. \ref{fig:xYc3_Upper_Bound_Accn}. The two curves through the Einstein fixed points parallel to the $Y$-axis denote where acceleration is zero. The $0.15 < x < 0.31$ region of the phase space is decelerating, and the $x < 0.15$ and $x > 0.31$ regions are accelerating. Therefore, bouncing models outside the CFS evolve with early- and late-time acceleration, with a decelerating phase in between, and are therefore the models of interest.  

\begin{figure}
\begin{center}
\includegraphics[width=80mm]{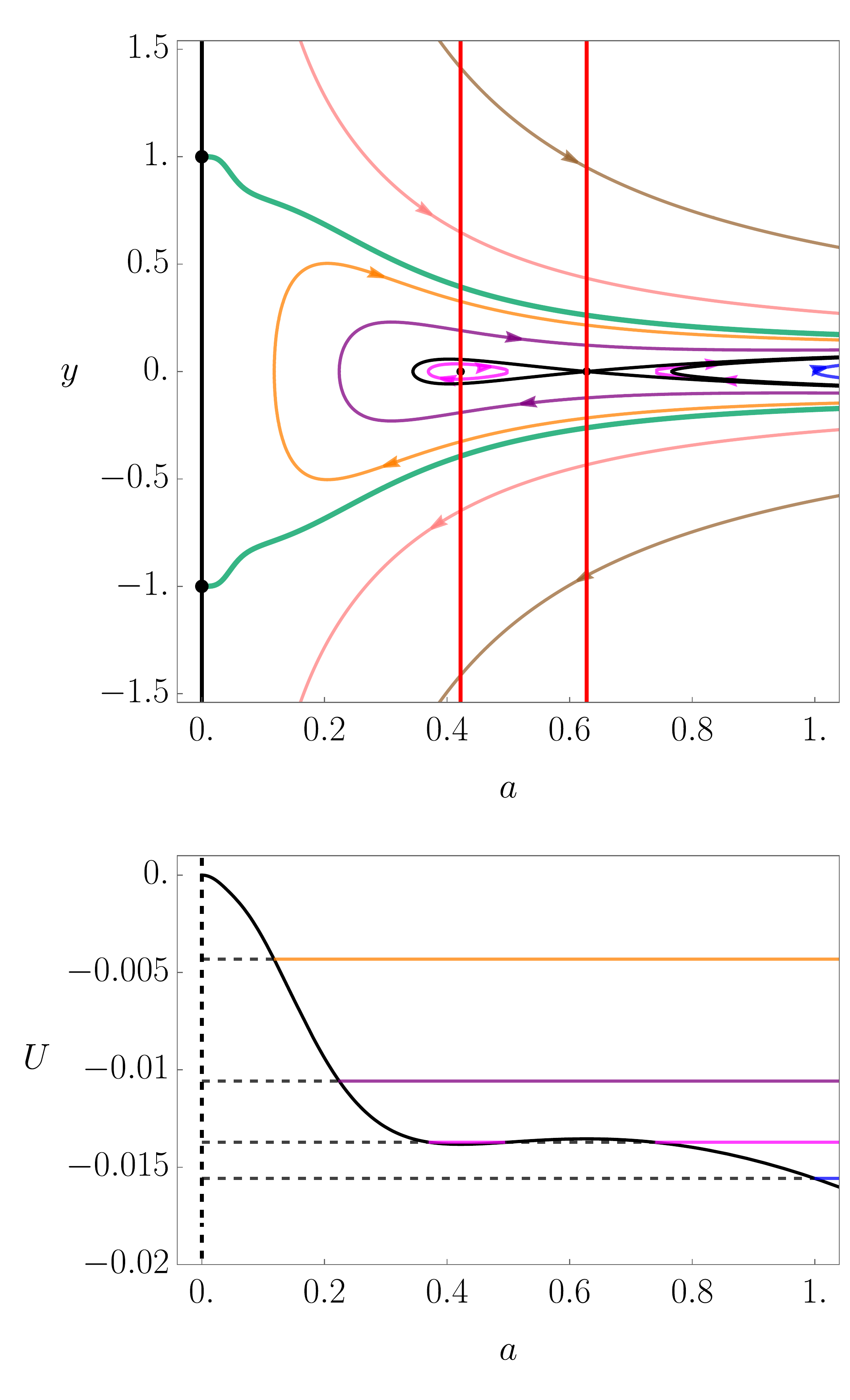}
\end{center}
\caption{The phase space (top panel) and corresponding potential (bottom panel, where now $U=U(a)$) for $0.25 < c_z < 11.0$ in terms of the scale factor $a$ and non-compactified Hubble expansion scalar $y$. This plot is equivalent to Fig. \ref{fig:xYc3_Upper_Bound_Accn}, with the boundaries separating the accelerating and decelerating regions shown in red. The region in between the two Einstein fixed points is decelerating, and the regions outside are accelerating.}
\label{fig:ayc3_Upper_Bound_Accn}
\end{figure}

Finally, using $a'=ay$ and $a''/a=y'+y^2$ in Eq. \eqref{eqn:accnupperbound} (where each variable now is a function of $a$), in Fig. \ref{fig:ayc3_Upper_Bound_Accn} we illustrate this model with the phase space for the scale factor $a$ (normalised to $a_o=1$ today)  and non-compactified Hubble expansion scalar $y$. Again, we see in this plot that at high energies, i.e. small $a$, all models have an accelerated phase, followed by a decelerated one and a final accelerated one at recent times, i.e.\ when $a \rightarrow 1$.

\subsection{A note on $\scriptR$} \label{sec:Script_R}

As we mentioned previously, in order to alleviate the old cosmological constant problem we would set $10^{-120} < \scriptR < 10^{-60}$, however purely for the purpose of illustrating the full range of dynamics in readable phase plots, we have set $\scriptR = 0.05$. If we had set $10^{-120} < \scriptR < 10^{-60}$, only one case for the dynamics would remain in each system. For both systems we have considered, we find from Eq. \eqref{eqn:c_r} and Eq. \eqref{eqn:cr_upper_bound} that for $10^{-120} < \scriptR < 10^{-60}$, we obtain $10^{16} < c_r < 10^{36}$. The effect of increasing radiation has qualitatively the same effect as increasing matter. 

For the first system of equations \eqref{CompactDarkMatterDE} - \eqref{CompactHubbleRateDE} where $z$ and $r$ are unbounded, we find the only dynamical case left, regardless of the value of $c_z$, is as in Fig. \ref{fig:xYc7}. The Einstein point would be pushed further towards $x = \scriptR$ and the closed models are mostly turn-around models which always decelerate. The bouncing models in this case are always accelerating, which is not the evolution we require for a realistic model.

For the second system of equations \eqref{CompactDarkEnergyDE}, \eqref{eqn:z'_upper_bound}, \eqref{eqn:r'_upper_bound} and \eqref{eqn:Y'_Upper_Limits} where $z$ and $r$ have an upper bound, we find that the only dynamical case is as in Fig. \ref{fig:xYc1_Upper_Bound}. All closed models bounce, however they are always accelerating, which again is not the evolution we require. 

As a final remark, we note that in order for a bouncing model to be theoretically robust and realistic, the value of $\scriptR$ would need to satisfy the bounds $10^{-120} < \scriptR < 10^{-60}$, producing a bounce with an early acceleration, followed by a decelerated era, and finally a late-time acceleration as in Fig. \ref{fig:xYc3_Upper_Bound_Accn}. The decelerated era should be a standard matter and radiation dominated phase, in order to satisfy observational constraints, but a contribution from the homogeneous dark energy remains a possibility.

\section{Conclusions} \label{sec:Conclusions}

In this paper, we have studied the dynamics of FLRW models containing dark matter, radiation and dark energy with a quadratic EoS. This is an extension of \cite{Ananda2006}, who studied a more general quadratic EoS in the high, low and full energy regimes without the inclusion of matter and radiation, and found bouncing and cyclic models were possible with certain combinations of parameters.
A quadratic EoS is the simplest nonlinear EoS and the qualitative analysis of the dynamics that it generates serves as guidance for more complicated nonlinear models; its study finds motivations in brane models, k-essence and loop quantum cosmology \cite{Shiromizu2000,Bridgman2002,Gergely2002,Langlois2003,Maartens2010,Vandersloot2005, Parisi2007,Scherrer2004,Giannakis2005} (see also \cite{Ananda2006,Ananda20062} and references therein).
We have restricted the EoS here so that the dark energy evolves between a high energy effective cosmological constant, which could be of order of the Planck energy, and a low energy effective cosmological constant close to the observed dark energy density today. Our aim was to investigate the effect of matter and radiation on the bouncing and cyclic models, to find whether these were still possible in this more realistic scenario. In particular, our focus was on all closed models avoiding singularities, instead having a bounce or cycles. Because the  EoS we consider is barotropic, naturally the evolution of the system is adiabatic, so that in phase space the expansion is a mirror image of the contraction, as in \cite{Ananda2006,Ananda20062,Bruni2022}. 

In Section \ref{sec:Dynamical_Eqns}, we have presented the system of equations for compactified variables, with the energy densities of dark matter and radiation satisfying the standard energy conservation equations, and as such are in principle unbounded and able to become infinite. In Section \ref{sec:Submanifolds}, we present the sub-manifolds of the system. In Section \ref{sec:Full_System}, we found one case of interest, shown in Fig. \ref{fig:xYc3_Accn}, where closed bouncing models can evolve with early and late time acceleration. However, we found that this bouncing behaviour is spoiled when matter and radiation become dominant for certain values of the initial conditions, and instead the system evolves with a turn-around between two singularities. In comparison to Fig. \ref{fig:Z0R02E} (with the same parameters but no matter and radiation), in which all closed models bounce, we concluded that these closed models could only be viable, i.e.\ always have a bounce independently of the initial conditions, if matter and radiation are only present at energies below the energy scale of the bounce. This would require a process such as reheating during the expansion phase after the bounce.

In light of this, we modelled the process by simply  introducing upper bounds on dark matter and radiation in Section \ref{sec:FullSystemUpperBound}, in order to avoid any models becoming singular. To keep things simple, with no need of extra parameters, we set this upper energy scale in line with the high energy effective cosmological constant of the dark energy component. In this case,   all closed models  have a bounce. The case of interest, shown in Fig.\ \ref{fig:xYc3_Upper_Bound_Accn}, admits bouncing models that evolve with early- and late-time acceleration which are not spoiled by the inclusion of matter and radiation, as they only appear at the same energy scale as the dark energy.

However, we noted that for the specific case of a quadratic EoS, this case of interest is not possible with a model that can alleviate the old cosmological constant problem \cite{Weinberg1989, Straumann1999, Ellis2012}. For a model where the two cosmological constants differ by 60 to 120 orders of magnitude, we found that the only case for the dynamics of the system is as in Fig. \ref{fig:xYc1_Upper_Bound}. All closed models bounce, however they always accelerate and so the evolution is undesirable. 

Overall, our qualitative analysis shows that a model with nonlinear dark energy and radiation and matter can be a realistic scenario like the one shown in Fig. \ref{fig:xYc3_Upper_Bound_Accn} and \ref{fig:ayc3_Upper_Bound_Accn}, with an early- and a late-time acceleration, where all closed models have a bounce, assuming that matter and radiation appear only after the bounce through a process such as reheating. However, while the simple specific model for dark energy considered here helps our understanding of precisely what the qualitative realistic scenario should be, it does not give the right quantitative behaviour. Learning from the present analysis, in future work we will consider models for dark energy, or for an interacting vacuum \cite{Bruni2022}, giving the qualitative behaviour of Fig. \ref{fig:xYc3_Upper_Bound_Accn} and \ref{fig:ayc3_Upper_Bound_Accn}, but with a realistic quantitative behaviour that will be worth constraining with data. Another interesting possibility that we leave for a future analysis will be to include a reheating phase through an interaction term between the dark energy and radiation components, with a possible role of dark matter.

\section*{Acknowledgements}

We thank Johannes Noller and David Wands for useful discussions and comments.
This work has been supported by UK STFC grants ST/S000550/1 and ST/T506345/1.

\onecolumngrid
\bibliography{refs}

\end{document}